\input harvmac

\noblackbox

\let\includefigures=\iftrue
\let\useblackboard=\iftrue
\newfam\black

%Figure Stuff
\includefigures
\message{If you do not have epsf.tex (to include figures),}
\message{change the option at the top of the tex file.}
\input epsf
\def\figin{\epsfcheck\figin}\def\figins{\epsfcheck\figins}
\def\epsfcheck{\ifx\epsfbox\UnDeFiNeD
\message{(NO epsf.tex, FIGURES WILL BE IGNORED)}
\gdef\figin##1{\vskip2in}\gdef\figins##1{\hskip.5in}% blank space instead
\else\message{(FIGURES WILL BE INCLUDED)}%
\gdef\figin##1{##1}\gdef\figins##1{##1}\fi}
\def\DefWarn#1{}
\def\figinsert{\goodbreak\midinsert}
\def\ifig#1#2#3{\DefWarn#1\xdef#1{Fig.~\the\figno}
\writedef{#1\leftbracket Fig.\noexpand~\the\figno}%
\figinsert\figin{\centerline{#3}}\medskip\centerline{\vbox{
\baselineskip12pt\advance\hsize by -1truein
\noindent\footnotefont{\bf Fig.~\the\figno:} #2}}
\bigskip\endinsert\global\advance\figno by1}
%%%
\else
\def\ifig#1#2#3{\xdef#1{Fig.~\the\figno}
\writedef{#1\leftbracket Fig.\noexpand~\the\figno}%
%\figinsert\figin{\centerline{#3}}\medskip
%\centerline{\vbox{\baselineskip12pt
%\advance\hsize by -1truein\noindent
%\footnotefont{\bf Fig.~\the\figno:} #2}}
%\bigskip\endinsert
\global\advance\figno by1} \fi

\def\doublefig#1#2#3#4{\DefWarn#1\xdef#1{Fig.~\the\figno}
\writedef{#1\leftbracket Fig.\noexpand~\the\figno}%
\figinsert\figin{\centerline{#3\hskip1.0cm#4}}\medskip\centerline{\vbox{
\baselineskip12pt\advance\hsize by -1truein
\noindent\footnotefont{\bf Fig.~\the\figno:} #2}}
\bigskip\endinsert\global\advance\figno by1}

%%BLACKBOARD FONT STUFF
\useblackboard
\message{If you do not have msbm (blackboard bold) fonts,}
\message{change the option at the top of the tex file.}
\font\blackboard=msbm10 scaled \magstep1 \font\blackboards=msbm7
\font\blackboardss=msbm5 \textfont\black=\blackboard
\scriptfont\black=\blackboards \scriptscriptfont\black=\blackboardss
\def\Bbb#1{{\fam\black\relax#1}}
\else
\def\Bbb{\bf}
\fi
% *************************************
%\draft
%

\def\yboxit#1#2{\vbox{\hrule height #1 \hbox{\vrule width #1
\vbox{#2}\vrule width #1 }\hrule height #1 }}
\def\fillbox#1{\hbox to #1{\vbox to #1{\vfil}\hfil}}
\def\ybox{{\lower 1.3pt \yboxit{0.4pt}{\fillbox{8pt}}\hskip-0.2pt}}
%
%
%%MATH MACROS
%Greek letters and their bars

%manifold and its universal cover

%More bars

%\def\l{\left}
%\def\r{\right}
\def\comments#1{}

\def\QH{\Bbb{H}}

\def\QR{\Bbb{R}}

\def\p{\partial}

\def\CN{{\cal N}}
%AEL

%AEL

\def\II{\relax{I\kern-.10em I}}

\def\IZ{\relax{\rm Z\kern-.34em Z}}
\def\IB{\relax{\rm I\kern-.18em B}}
\def\IC{{\relax\hbox{$\inbar\kern-.3em{\rm C}$}}}
\def\ID{\relax{\rm I\kern-.18em D}}
\def\IE{\relax{\rm I\kern-.18em E}}
\def\IF{\relax{\rm I\kern-.18em F}}
\def\IG{\relax\hbox{$\inbar\kern-.3em{\rm G}$}}
\def\IGa{\relax\hbox{${\rm I}\kern-.18em\Gamma$}}
\def\IH{\relax{\rm I\kern-.18em H}}
\def\II{\relax{\rm I\kern-.18em I}}
\def\IK{\relax{\rm I\kern-.18em K}}
\def\IP{\relax{\rm I\kern-.18em P}}
%\def\IX{\relax{\rm X\kern-.01em X}}
%this doesn't work

%

\def\inbar{\,\vrule height1.5ex width.4pt depth0pt}

\def\p{\partial}

\def\IR{\relax{\rm I\kern-.18em R}}

\def\simgt{\hskip0.05in\relax{
\raise3.0pt\hbox{ $>$ {\lower5.0pt\hbox{\kern-1.05em $\sim$}} }}
\hskip0.05in}

%

 % for now

%

\def\lp10{\ell_p^{10}}
\def\lp11{\ell_p^{11}}
\def\R11{R_{11}}

\def\frac#1#2{{#1 \over #2}}

%identity operator from doyon-fonseca

%% from the topological vertex paper

%%                              TABLEAUX.TEX
%%      This  macro file is for producing a ``Young Tableau'' which is
%%      an array of little squares sometimes used in mathematical physics.
%%      For instance, the command $\tableau{6 3 2}$ will produce a tableau
%%      with 6 squares in the top row, 3 in the next, and 2 in the last.
%%                                  OOOOOO
%%      This tableau will look like OOO    but made of squares instead of O's.
%%                                  OO
%%      Any number of rows may be present, each having a nonzero number of
%%      squares.
%%
%%      A tableau is math mode material, so use $ or $$ to enclose it.
%%
%%      The size and line-thickness of the little boxes are controlled by the
%%      dimension parameters --
%%              \tableauside=1.0ex              %(size)
%%              \tableaurule=0.4pt              %(line-thickness)
%%      Change them if you want.
%%
%%                                                      -- Doug Eardley 9/19/8%%
%%
\newdimen\tableauside\tableauside=1.0ex
\newdimen\tableaurule\tableaurule=0.4pt
\newdimen\tableaustep
\def\phantomhrule#1{\hbox{\vbox to0pt{\hrule height\tableaurule width#1\vss}}}
\def\phantomvrule#1{\vbox{\hbox to0pt{\vrule width\tableaurule height#1\hss}}}
\def\sqr{\vbox{%
  \phantomhrule\tableaustep
  \hbox{\phantomvrule\tableaustep\kern\tableaustep\phantomvrule\tableaustep}%
  \hbox{\vbox{\phantomhrule\tableauside}\kern-\tableaurule}}}
\def\squares#1{\hbox{\count0=#1\noindent\loop\sqr
  \advance\count0 by-1 \ifnum\count0>0\repeat}}
\def\tableau#1{\vcenter{\offinterlineskip
  \tableaustep=\tableauside\advance\tableaustep by-\tableaurule
  \kern\normallineskip\hbox
    {\kern\normallineskip\vbox
      {\gettableau#1 0 }%
     \kern\normallineskip\kern\tableaurule}%
  \kern\normallineskip\kern\tableaurule}}
\def\gettableau#1 {\ifnum#1=0\let\next=\null\else
  \squares{#1}\let\next=\gettableau\fi\next}

\tableauside=1.0ex \tableaurule=0.4pt

%% from shiraz

 %
 %       \eqn\label{a+b=c}       gives displayed equation, numbered
 %                               consecutively within sections.
%     \eqnn and \eqna define labels in advance (of eqalign?)
 %
 \def\eqnn#1{\xdef #1{(\secsym\the\meqno)}\writedef{#1\leftbracket#1}%
 \global\advance\meqno by1\wrlabeL#1}
 \def\eqna#1{\xdef #1##1{\hbox{$(\secsym\the\meqno##1)$}}
 \writedef{#1\numbersign1\leftbracket#1{\numbersign1}}%
 \global\advance\meqno by1\wrlabeL{#1$\{\}$}}
 \def\eqn#1#2{\xdef #1{(\secsym\the\meqno)}\writedef{#1\leftbracket#1}%
 \global\advance\meqno by1$$#2\eqno#1\eqlabeL#1$$}

\global\newcount\itemno \global\itemno=0

%\bigbreak}
%\bigskip\noindent}
\def\itemaut#1{\global\advance\itemno by1\noindent\item{\the\itemno.}#1}

%\itemized
%\itemaut{First this.}
%\itemaut{Then that.}

%%More macros
\def\hV{{\hat V}}
\def\tp{t_{p}}

\def\rp{r_{p}}
\def\tdt{\tilde{t}}
\def\tir{\tilde{r}}

%%ENGLISH MACROS

\hyphenation{Di-men-sion-al}

%%REFERENCING MACROS

%%

%\EmparanGF
\lref\EmparanGF{
  R.~Emparan,
  ``AdS/CFT duals of topological black holes and the entropy of  zero-energy
  states,''
  JHEP {\bf 9906}, 036 (1999)
  [arXiv:hep-th/9906040].
  %%CITATION = JHEPA,9906,036;%%
}

%\GreenTR
\lref\GreenTR{
  D.~Green, A.~Lawrence, J.~McGreevy, D.~R.~Morrison and E.~Silverstein,
  ``Dimensional Duality,''
  Phys.\ Rev.\  D {\bf 76}, 066004 (2007)
  [arXiv:0705.0550 [hep-th]].
  %%CITATION = PHRVA,D76,066004;%%
}

%\HertogRZ
\lref\HertogRZ{
  T.~Hertog and G.~T.~Horowitz,
  ``Towards a big crunch dual,''
  JHEP {\bf 0407}, 073 (2004)
  [arXiv:hep-th/0406134];
  %%CITATION = JHEPA,0407,073;%%
T.~Hertog and G.~T.~Horowitz,
  ``Holographic description of AdS cosmologies,''
  JHEP {\bf 0504}, 005 (2005)
  [arXiv:hep-th/0503071];
  %%CITATION = JHEPA,0504,005;%%
%\cite{Hertog:2004ns}
%\bibitem{Hertog:2004ns}
  T.~Hertog and G.~T.~Horowitz,
  ``Designer gravity and field theory effective potentials,''
  Phys.\ Rev.\ Lett.\  {\bf 94}, 221301 (2005)
  [arXiv:hep-th/0412169].
  %%CITATION = PRLTA,94,221301;%%
}

%\CrapsCH
\lref\CrapsCH{
  B.~Craps, T.~Hertog and N.~Turok,
  ``Quantum Resolution of Cosmological Singularities using AdS/CFT,''
  arXiv:0712.4180 [hep-th].
  %%CITATION = ARXIV:0712.4180;%%
}

%\KrausIV
\lref\KrausIV{
  P.~Kraus, H.~Ooguri and S.~Shenker,
  ``Inside the horizon with AdS/CFT,''
  Phys.\ Rev.\  D {\bf 67}, 124022 (2003)
  [arXiv:hep-th/0212277];
  %%CITATION = PHRVA,D67,124022;%%
%\FidkowskiNF
%\lref\FidkowskiNF{
  L.~Fidkowski, V.~Hubeny, M.~Kleban and S.~Shenker,
  ``The black hole singularity in AdS/CFT,''
  JHEP {\bf 0402}, 014 (2004)
  [arXiv:hep-th/0306170].
  %%CITATION = JHEPA,0402,014;%%
%}
}

%\AwadFJ
\lref\AwadFJ{
  A.~Awad, S.~R.~Das, K.~Narayan and S.~P.~Trivedi,
  ``Gauge Theory Duals of Cosmological Backgrounds and their Energy Momentum
  Tensors,''
  Phys.\ Rev.\  D {\bf 77}, 046008 (2008)
  [arXiv:0711.2994 [hep-th]];
  %%CITATION = PHRVA,D77,046008;%%
%\DasPW
%\lref\DasPW{
  S.~R.~Das, J.~Michelson, K.~Narayan and S.~P.~Trivedi,
  %``Cosmologies with null singularities and their gauge theory duals,''
  Phys.\ Rev.\  D {\bf 75}, 026002 (2007)
  [arXiv:hep-th/0610053];
  %%CITATION = PHRVA,D75,026002;%%
%\DasDZ
%\lref\DasDZ{
  S.~R.~Das, J.~Michelson, K.~Narayan and S.~P.~Trivedi,
  ``Time dependent cosmologies and their duals,''
  Phys.\ Rev.\  D {\bf 74}, 026002 (2006)
  [arXiv:hep-th/0602107].
  %%CITATION = PHRVA,D74,026002;%%
}

%\DouglasYP
\lref\DouglasYP{
  M.~R.~Douglas, D.~N.~Kabat, P.~Pouliot and S.~H.~Shenker,
  ``D-branes and short distances in string theory,''
  Nucl.\ Phys.\  B {\bf 485}, 85 (1997)
  [arXiv:hep-th/9608024].
  %%CITATION = NUPHA,B485,85;%%
}

%\KofmanYC
\lref\KofmanYC{
  L.~Kofman, A.~Linde, X.~Liu, A.~Maloney, L.~McAllister and E.~Silverstein,
  ``Beauty is attractive: Moduli trapping at enhanced symmetry points,''
  JHEP {\bf 0405}, 030 (2004)
  [arXiv:hep-th/0403001].
  %%CITATION = JHEPA,0405,030;%%
}

%\SilversteinHF
\lref\SilversteinHF{
  E.~Silverstein and D.~Tong,
  ``Scalar speed limits and cosmology: Acceleration from D-cceleration,''
  Phys.\ Rev.\  D {\bf 70}, 103505 (2004)
  [arXiv:hep-th/0310221].
  %%CITATION = PHRVA,D70,103505;%%
}

%\deBoerXF
\lref\deBoerXF{
  J.~de Boer, E.~P.~Verlinde and H.~L.~Verlinde,
  ``On the holographic renormalization group,''
  JHEP {\bf 0008}, 003 (2000)
  [arXiv:hep-th/9912012].
  %%CITATION = JHEPA,0008,003;%%
}

%\deBoerCZ
\lref\deBoerCZ{
  J.~de Boer,
  ``The holographic renormalization group,''
  Fortsch.\ Phys.\  {\bf 49}, 339 (2001)
  [arXiv:hep-th/0101026].
  %%CITATION = FPYKA,49,339;%%
}

%\EmparanGF
\lref\EmparanGF{
  R.~Emparan,
  ``AdS/CFT duals of topological black holes and the entropy of  zero-energy
  states,''
  JHEP {\bf 9906}, 036 (1999)
  [arXiv:hep-th/9906040].
  %%CITATION = JHEPA,9906,036;%%
}

%\HertogRZ
\lref\HertogRZ{
  T.~Hertog and G.~T.~Horowitz,
  ``Towards a big crunch dual,''
  JHEP {\bf 0407}, 073 (2004)
  [arXiv:hep-th/0406134];
  %%CITATION = JHEPA,0407,073;%%
T.~Hertog and G.~T.~Horowitz,
  ``Holographic description of AdS cosmologies,''
  JHEP {\bf 0504}, 005 (2005)
  [arXiv:hep-th/0503071].
  %%CITATION = JHEPA,0504,005;%%
%\cite{Hertog:2004ns}
}

%\CrapsCH
\lref\CrapsCH{
  B.~Craps, T.~Hertog and N.~Turok,
  ``Quantum Resolution of Cosmological Singularities using AdS/CFT,''
  arXiv:0712.4180 [hep-th].
  %%CITATION = ARXIV:0712.4180;%%
}

%\DvaliTQ
\lref\Othertrap{
  G.~R.~Dvali,
  ``Infrared hierarchy, thermal brane inflation and superstrings as  superheavy
  dark matter,''
  Phys.\ Lett.\  B {\bf 459}, 489 (1999)
  [arXiv:hep-ph/9905204];
  %%CITATION = PHLTA,B459,489;%%
%\cite{Traschen:1990sw}
%\bibitem{Traschen:1990sw}
  J.~H.~Traschen and R.~H.~Brandenberger,
  ``Particle production during out-of-equilibrium phase transitions,''
  Phys.\ Rev.\  D {\bf 42}, 2491 (1990).
  %%CITATION = PHRVA,D42,2491;%%
}

%\AwadFJ
\lref\AwadFJ{
  A.~Awad, S.~R.~Das, K.~Narayan and S.~P.~Trivedi,
  ``Gauge Theory Duals of Cosmological Backgrounds and their Energy Momentum
  Tensors,''
  Phys.\ Rev.\  D {\bf 77}, 046008 (2008)
  [arXiv:0711.2994 [hep-th]];
  %%CITATION = PHRVA,D77,046008;%%
%\DasPW
%\lref\DasPW{
  S.~R.~Das, J.~Michelson, K.~Narayan and S.~P.~Trivedi,
  %``Cosmologies with null singularities and their gauge theory duals,''
  Phys.\ Rev.\  D {\bf 75}, 026002 (2007)
  [arXiv:hep-th/0610053];
  %%CITATION = PHRVA,D75,026002;%%
%\DasDZ
%\lref\DasDZ{
  S.~R.~Das, J.~Michelson, K.~Narayan and S.~P.~Trivedi,
  ``Time dependent cosmologies and their duals,''
  Phys.\ Rev.\  D {\bf 74}, 026002 (2006)
  [arXiv:hep-th/0602107].
  %%CITATION = PHRVA,D74,026002;%%
}

%\DouglasYP
\lref\DouglasYP{
  M.~R.~Douglas, D.~N.~Kabat, P.~Pouliot and S.~H.~Shenker,
  ``D-branes and short distances in string theory,''
  Nucl.\ Phys.\  B {\bf 485}, 85 (1997)
  [arXiv:hep-th/9608024].
  %%CITATION = NUPHA,B485,85;%%
}

%\KofmanYC
\lref\KofmanYC{
  L.~Kofman, A.~Linde, X.~Liu, A.~Maloney, L.~McAllister and E.~Silverstein,
  ``Beauty is attractive: Moduli trapping at enhanced symmetry points,''
  JHEP {\bf 0405}, 030 (2004)
  [arXiv:hep-th/0403001].
  %%CITATION = JHEPA,0405,030;%%
}

%\SilversteinHF
\lref\SilversteinHF{
  E.~Silverstein and D.~Tong,
  ``Scalar speed limits and cosmology: Acceleration from D-cceleration,''
  Phys.\ Rev.\  D {\bf 70}, 103505 (2004)
  [arXiv:hep-th/0310221].
  %%CITATION = PHRVA,D70,103505;%%
}

%\deBoerXF
\lref\deBoerXF{
  J.~de Boer, E.~P.~Verlinde and H.~L.~Verlinde,
  ``On the holographic renormalization group,''
  JHEP {\bf 0008}, 003 (2000)
  [arXiv:hep-th/9912012].
  %%CITATION = JHEPA,0008,003;%%
}

%\HamiltonFH
\lref\HamiltonFH{
  A.~Hamilton, D.~N.~Kabat, G.~Lifschytz and D.~A.~Lowe,
  ``Local bulk operators in AdS/CFT: A holographic description of the black
  hole interior,''
  Phys.\ Rev.\  D {\bf 75}, 106001 (2007)
  [Erratum-ibid.\  D {\bf 75}, 129902 (2007)]
  [arXiv:hep-th/0612053].
  %%CITATION = PHRVA,D75,106001;%%
}

%\MarolfMF
\lref\MarolfMF{
  D.~Marolf,
  ``Unitarity and Holography in Gravitational Physics,''
  arXiv:0808.2842 [gr-qc].
  %%CITATION = ARXIV:0808.2842;%%
}

%\deBoerCZ
\lref\deBoerCZ{
  J.~de Boer,
  ``The holographic renormalization group,''
  Fortsch.\ Phys.\  {\bf 49}, 339 (2001)
  [arXiv:hep-th/0101026].
  %%CITATION = FPYKA,49,339;%%
}

%\AsplundXD
\lref\AsplundXD{
  C.~T.~Asplund and D.~Berenstein,
  ``Small AdS black holes from SYM,''
  arXiv:0809.0712 [hep-th].
  %%CITATION = ARXIV:0809.0712;%%
}
 %\HorowitzMR
\lref\HorowitzMR{
  G.~T.~Horowitz and E.~Silverstein,
  ``The inside story: Quasilocal tachyons and black holes,''
  Phys.\ Rev.\  D {\bf 73}, 064016 (2006)
  [arXiv:hep-th/0601032].
  %%CITATION = PHRVA,D73,064016;%%
  }
%\MaldacenaRE
\lref\MaldacenaRE{
  J.~M.~Maldacena,
  ``The large N limit of superconformal field theories and supergravity,''
  Adv.\ Theor.\ Math.\ Phys.\  {\bf 2}, 231 (1998)
  [Int.\ J.\ Theor.\ Phys.\  {\bf 38}, 1113 (1999)]
  [arXiv:hep-th/9711200].
  %%CITATION = IJTPB,38,1113;%%
}
%\KiritsisTX
\lref\KiritsisTX{
  E.~Kiritsis,
  ``Supergravity, D-brane probes and thermal super Yang-Mills:  A comparison,''
  JHEP {\bf 9910}, 010 (1999)
  [arXiv:hep-th/9906206].
  %%CITATION = JHEPA,9910,010;%%
}
%\JevickiQS
\lref\JevickiQS{
  A.~Jevicki, Y.~Kazama and T.~Yoneya,
  ``Quantum metamorphosis of conformal transformation in D3-brane  Yang-Mills
  theory,''
  Phys.\ Rev.\ Lett.\  {\bf 81}, 5072 (1998)
  [arXiv:hep-th/9808039].
  %%CITATION = PRLTA,81,5072;%%
}
%\MaldacenaNX
\lref\MaldacenaNX{
  J.~M.~Maldacena,
  ``Probing near extremal black holes with D-branes,''
  Phys.\ Rev.\  D {\bf 57}, 3736 (1998)
  [arXiv:hep-th/9705053].
  %%CITATION = PHRVA,D57,3736;%%
}
%\MaldacenaKR
\lref\MaldacenaKR{
  J.~M.~Maldacena,
  ``Eternal black holes in Anti-de-Sitter,''
  JHEP {\bf 0304}, 021 (2003)
  [arXiv:hep-th/0106112].
  %%CITATION = JHEPA,0304,021;%%
}
%\HorowitzXK
\lref\HorowitzXK{
  G.~T.~Horowitz and D.~Marolf,
  ``A new approach to string cosmology,''
  JHEP {\bf 9807}, 014 (1998)
  [arXiv:hep-th/9805207].
  %%CITATION = JHEPA,9807,014;%%
}
%\BalasubramanianDE
\lref\BalasubramanianDE{
  V.~Balasubramanian, P.~Kraus, A.~E.~Lawrence and S.~P.~Trivedi,
  ``Holographic probes of anti-de Sitter space-times,''
  Phys.\ Rev.\  D {\bf 59}, 104021 (1999)
  [arXiv:hep-th/9808017].
  %%CITATION = PHRVA,D59,104021;%%
}
%\CarneirodaCunhaJF
\lref\CarneirodaCunhaJF{
  B.~G.~Carneiro da Cunha,
  ``Inflation and holography in string theory,''
  Phys.\ Rev.\  D {\bf 65}, 026001 (2002)
  [arXiv:hep-th/0105219].
  %%CITATION = PHRVA,D65,026001;%%
}
%\JohnsonQT
\lref\JohnsonQT{
  C.~V.~Johnson, A.~W.~Peet and J.~Polchinski,
  ``Gauge theory and the excision of repulson singularities,''
  Phys.\ Rev.\  D {\bf 61}, 086001 (2000)
  [arXiv:hep-th/9911161].
  %%CITATION = PHRVA,D61,086001;%%
}
%\CrapsWD
\lref\CrapsWD{
  B.~Craps, S.~Sethi and E.~P.~Verlinde,
  ``A Matrix Big Bang,''
  JHEP {\bf 0510}, 005 (2005)
  [arXiv:hep-th/0506180];
  %%CITATION = JHEPA,0510,005;%%
  %\RobbinsUA
%\lref\RobbinsUA{
  D.~Robbins and S.~Sethi,
  ``A matrix model for the null-brane,''
  JHEP {\bf 0602}, 052 (2006)
  [arXiv:hep-th/0509204].
  %%CITATION = JHEPA,0602,052;%%
  %\CrapsXQ
%\lref\CrapsXQ{
  B.~Craps, A.~Rajaraman and S.~Sethi,
  ``Effective dynamics of the matrix big bang,''
  Phys.\ Rev.\  D {\bf 73}, 106005 (2006)
  [arXiv:hep-th/0601062];
  %\MartinecAK
%\lref\MartinecAK{
  E.~J.~Martinec, D.~Robbins and S.~Sethi,
  ``Toward the end of time,''
  JHEP {\bf 0608}, 025 (2006)
  [arXiv:hep-th/0603104].
  %%CITATION = JHEPA,0608,025;%%
%}
  %%CITATION = PHRVA,D73,106005;%%
%}
%}
}

%\DasDZ
\lref\DasDZ{
  S.~R.~Das, J.~Michelson, K.~Narayan and S.~P.~Trivedi,
  %``Time dependent cosmologies and their duals,''
  Phys.\ Rev.\  D {\bf 74}, 026002 (2006)
  [arXiv:hep-th/0602107];
  %%CITATION = PHRVA,D74,026002;%%
 %\DasPW
%\lref\DasPW{
  S.~R.~Das, J.~Michelson, K.~Narayan and S.~P.~Trivedi,
  ``Cosmologies with Null Singularities and their Gauge Theory Duals,''
  Phys.\ Rev.\  D {\bf 75}, 026002 (2007)
  [arXiv:hep-th/0610053];
  %%CITATION = PHRVA,D75,026002;%%
%}
%\AwadFJ
%\lref\AwadFJ{
  A.~Awad, S.~R.~Das, K.~Narayan and S.~P.~Trivedi,
  ``Gauge Theory Duals of Cosmological Backgrounds and their Energy Momentum
  Tensors,''
  Phys.\ Rev.\  D {\bf 77}, 046008 (2008)
  [arXiv:0711.2994 [hep-th]];
  %%CITATION = PHRVA,D77,046008;%%
%}
%\AwadJF
%\lref\AwadJF{
  A.~Awad, S.~R.~Das, S.~Nampuri, K.~Narayan and S.~P.~Trivedi,
  ``Gauge Theories with Time Dependent Couplings and their Cosmological
  Duals,''
  arXiv:0807.1517 [hep-th].
  %%CITATION = ARXIV:0807.1517;%%
%}
}

%\KabatVC
\lref\KabatVC{
  D.~N.~Kabat and G.~Lifschytz,
  ``Tachyons and black hole horizons in gauge theory,''
  JHEP {\bf 9812}, 002 (1998)
  [arXiv:hep-th/9806214];
    %%CITATION = JHEPA,9905,005;%%
%}
%\KabatYQ
%\lref\KabatYQ{
  D.~N.~Kabat and G.~Lifschytz,
  ``Gauge theory origins of supergravity causal structure,''
  JHEP {\bf 9905}, 005 (1999)
  [arXiv:hep-th/9902073];
  %%CITATION = JHEPA,9812,002;%%
  %\IizukaCW
%\lref\IizukaCW{
  N.~Iizuka, D.~N.~Kabat, G.~Lifschytz and D.~A.~Lowe,
  ``Probing black holes in non-perturbative gauge theory,''
  Phys.\ Rev.\  D {\bf 65}, 024012 (2002)
  [arXiv:hep-th/0108006].
  %%CITATION = PHRVA,D65,024012;%%
%}
}

%\CornishHZ
\lref\CornishHZ{
  N.~J.~Cornish, D.~Spergel and G.~Starkman,
  ``Can COBE see the shape of the universe?,''
  Phys.\ Rev.\  D {\bf 57}, 5982 (1998)
  [arXiv:astro-ph/9708225].
  %%CITATION = PHRVA,D57,5982;%%
}
%\MullerPE
\lref\MullerPE{
  D.~Muller, H.~V.~Fagundes and R.~Opher,
  ``Casimir energy in a small volume multiply connected static hyperbolic
  pre-inflationary universe,''
  Phys.\ Rev.\  D {\bf 63}, 123508 (2001)
  [arXiv:gr-qc/0103014].
  %%CITATION = PHRVA,D63,123508;%%
}

\Title{\vbox{\baselineskip12pt\hbox{NSF-KITP-09-53}
\hbox{SLAC-PUB-13584}\hbox{SU-ITP-09/16}\hbox{BRX-TH-607}}} {\vbox{ \centerline{}
%\medskip
\centerline{Insightful D-branes}}
%
%Insight
%
%Insightful 3-branes
%
%D-brane Insight into Black Holes
%
%
%
}
\centerline{Gary Horowitz$^{1}$, Albion Lawrence$^{2,4}$, and Eva Silverstein$^{3,4}$}
\medskip
%\centerline{$^2${\it
%Kavli Institute for Theoretical Physics, University of California,
%Santa Barbara, CA 93106-4030}}
\centerline{$^1$\it{Department of Physics, University of California, Santa Barbara, CA 93106}}
\centerline{$^2${\it Brandeis Theory Group, Brandeis University, MS 057, PO Box 549110, Waltham, MA 02454}}
\centerline{$^3${\it SLAC and Department of Physics,
Stanford University, Stanford, CA 94305-4060}}
\centerline{$^4${\it Kavli Institute for Theoretical Physics, University of California, Santa Barbara,
CA 93106}}
%\medskip \bigskip

\vskip.4in
We study a simple model of a black hole in AdS and obtain a holographic description of the region inside the horizon. A key role is played by the dynamics of the scalar fields in the dual gauge theory. This leads to a proposal for a dual description of D-branes falling through the horizon of any AdS black hole. The proposal uses a field-dependent time reparameterization in the field theory. We relate
this reparametrization to various gauge invariances of the theory.  Finally, we speculate
on information loss and the black hole singularity in this context.

%\draftmode

\Date{April 2009}

\newsec{Introduction}

A longstanding problem in gauge/gravity duality is to describe the region inside a black hole horizon in terms of the dual gauge theory. There have been various interesting approaches to this problem \refs{\KrausIV,\HamiltonFH,\MarolfMF}  but none have directly addressed the question of how to describe the observations of an infalling observer. We will present a proposal for how to do this, using the dynamics of rolling scalar fields in the gauge theory and their manifestation in terms of D-brane probes on the gravity side.

Consider the patch of $AdS_5$, with metric
\eqn\poincmet{ds^2 = {r_p^2\over \ell^2} (-dt_p^2 + t_p^2 d \sigma^2) + {\ell^2\over r_p^2} dr_p^2.}
Here we have started with Poincar\'e coordinates, but replaced the constant-$r_p$
Minkowski space slices $\QR^{3,1}$ with a patch of Minkowski space consisting of
the backward light cone of a point; this patch is described by Milne coordinates with spatial slices equal to hyperbolic 3-space $\QH_3$, with constant curvature
metric $d\sigma^2$. The dual gauge theory describing this
lives on the geometry
\eqn\ccgeom{
	ds^2 = - dt_p^2 + t_p^2 d\sigma^2\
}
or equivalently on the static cylinder geometry
\eqn\staticcyl{
	ds^2 = -d\eta^2+ \ell^2d\sigma^2
}
obtained from \ccgeom\ by a conformal transformation that is valid away from $t_p=0$.

Placing a D3-brane at constant $r_p$ is dual to the gauge theory with a scalar field VEV turned on breaking its $U(N)$ gauge symmetry down to $U(N-1)\times U(1)$.  The radial collective coordinate of the D3-brane, which corresponds to the VEV of the scalar field eigenvalue, is governed at low energies by the DBI action.
%{\tt [I think this is too complicated to put in the intro. Do we really need the explicit form here?]}
%
%\eqn\poincDBI{S_{DBIP} = -\int d^4 x {r_p^4 t_p^3\over{\ell^4 g_s\alpha'^2}} \left(\sqrt{1-\frac{\ell^4 (\dot r_p^2-{\ell^2\over t^2}(\nabla r_p)^2)}{r_p^4}}-1\right).}
%
This is obtained by integrating out the other degrees of freedom in the gauge theory.\foot{A similar action describes the low energy dynamics of the full set of low energy degrees of freedom on the D3-brane.
One can also generalize to a stack of $n\ll N$ D3-branes with interacting degrees of freedom; a fair
amount is known about this``non-abelian DBI" action.}

Compactifying the hyperbolic space, by orbifolding $\QH_3$ by a discrete group of
isometries $\Gamma$, leads to a spacetime with a singularity at $t_p=0$.  As we will review in \S2, this spacetime is in fact a static black hole \EmparanGF\ whose event horizon is shown in figure 1a.  (This is a higher dimensional generalization of the BTZ black hole in three dimensions.)\foot{Other
singularities in spacetimes with holographic duals have been described in \refs{\CrapsWD,\DasDZ}.
These correspond to gauge theories with dimensionful or time-dependent couplings, so there
is no static, non-singular presentation of the gauge theory as our system has.}

\ifig\gravpics{A black hole can be obtained from the Poincare patch by compactifying along hyperbolic spatial slices. The (green) dash-dot line denotes a D3-brane probe. (a) The (blue) dashed lines denote spatial slices in Poincar\'e coordinates (b)  The (blue) dashed lines are spatial slices in Schwarzschild coordinates.} {\epsfxsize3.0in\epsfbox{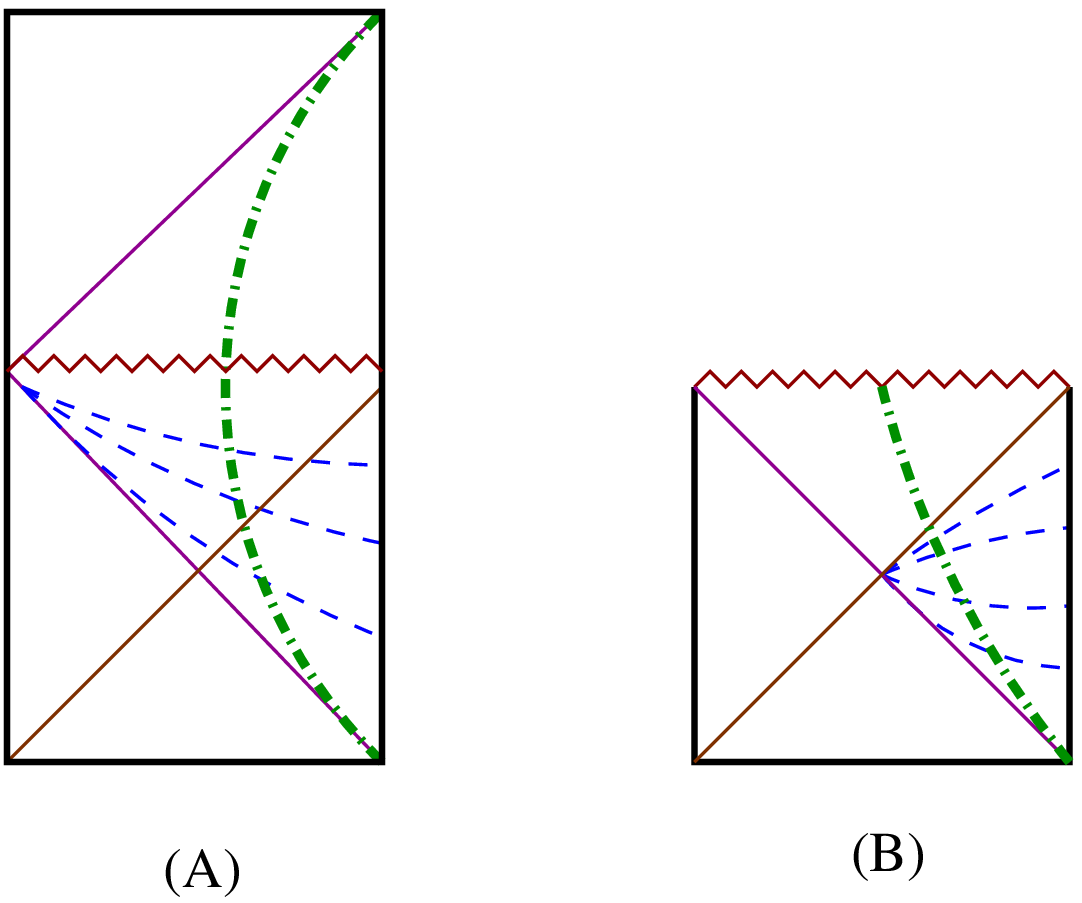}}

In the figure, we also indicate the trajectory of the D3-brane at constant $r_p$.  Well before approaching the singularity, while its dynamics is still described to a very good approximation by the DBI action,
%\poincDBI,
this D3-brane falls through the horizon of the black hole.    An infalling observer on the D3-brane is thus described in the Yang-Mills theory by the low energy effective action for the scalar field $r_p$ and its superpartners. We will show that these variables remain classical through the horizon, providing a simple description of an observer falling inside the black hole.

An observer outside the black hole, on the other hand, can be described on the gravity side by standard Schwarzschild coordinates \refs{\EmparanGF}, with metric:
\eqn\sccord{
	ds^2 = - \left(\frac{r^2}{\ell^2} - 1\right) dt^2 + \frac{dr^2}{\left(\frac{r^2}{\ell^2} - 1\right)}
		+ r^2 d\sigma^2\ .
}
The equal-$t$ time slices are depicted in figure 1b.  As we will describe below,
the field theory duals of the two descriptions are related by a conformal transformation to a static cylinder \staticcyl\
combined with a field-dependent time reparameterization. This second set of gauge theory variables describes the black hole in terms of a thermal field theory. The combined transformation produces a DBI action in Schwarzschild coordinates.  This effective action breaks down near the horizon because of the coordinate singularity there.  It takes forever to reach the horizon in Schwarzschild time $t$, and as we will discuss further below the fluctuations about the classical solution grow large in these variables.  These features reflect the thermalization of initial excitations in the gauge theory.

Nonetheless, both sets of variables descend from the Yang-Mills description, as we will describe in
\S3, and so one should be able to map phenomena seen in one set of variables to phenomena in another.  We will describe and compare the physics in different
frames in \S3, and construct the transformation of quantum variables
between these frames in \S4.
Given these results, we can begin to attack some well-known problems in black hole physics.

The black hole can be formed naturally from a collapsing shell of $N$ D3-branes, and eventually decays as we will discuss in \S5.  The description of the black hole interior provided by the scalar field dynamics may help us gain insight into the problem of information loss.
%Our proposal is motivated by a simple example of matter collapsing to form a black hole which can be given a complete %dual description (including the region inside the horizon). The example involves a shell of D3-branes, each wrapped on %a negatively curved surface, which collapses to form a black hole with negatively curved horizon.
Although we began with hyperbolic black holes derived in the Poincar\'e patch of $AdS_5$,  our proposal can be generalized to apply to other types of black holes as well.

%The basic idea is simple. The standard description of the gauge theory on the boundary is believed to describe %evolution in the bulk in a Schwarzschild-like time coordinate $t$ which goes bad at the horizon. To obtain a time %coordinate which is nonsingular at the horizon,   one typically has to shift $t$ by a function of the radial %coordinate. Since the scalars in the gauge theory can be associated with the radial coordinate in the bulk (e.g., in %describing the location of D-branes), to implement the coordinate transformation in the bulk we need to introduce %field-dependent time reparameterizations in the gauge theory. We will describe what this means and how to implement %it.

Having found a way to follow matter falling inside the horizon, one can now ask what happens near the black hole singularity. At present our discussion is necessarily qualitative since both the bulk and the boundary theory become strongly coupled. Nevertheless, an interesting picture emerges, which we also discuss in \S5. The physics near the singularity is associated with a rolling scalar field, just like earlier discussions of singularities in AdS \refs{\HertogRZ,\CrapsCH}. However, there are several crucial differences. In these papers, the bulk singularities only arose after the  field theory was modified with a multi-trace operator which produced a potential unbounded from below.  The Hamiltonian was ill defined without some prescription for a self adjoint extension, and the singularity was associated with the scalar field rolling off to infinity; that is, with ultraviolet physics.  In the present case, although we again have a potential unbounded from below, there is no need for multi-trace operators or self adjoint extensions. Furthermore, in our class of examples, the location of the would-be singularity is at the origin in scalar field space rather than at infinity, a regime governed by infrared physics, specifically the corresponding effects of the extra light states at the origin along the lines of \refs{\DouglasYP,\SilversteinHF,\KofmanYC}.\foot{See \AsplundXD\
for a recent discussion of these effects in black hole physics in AdS/CFT, and \Othertrap\ for previous discussions of related effects.}  In particular, near the singularity there is no local description of the physics in terms of scalar field eigenvalues, since the full set of $U(N)$ adjoint matrix degrees of freedom participates in the dynamics.

We will present some further speculations and directions for future work in \S6.

%%%%%%
\newsec{Hyperbolic black holes}

The metric \poincmet\ has a Killing field which is timelike near infinity and spacelike near the singularity. To make this manifest, we introduce new coordinates
\eqn\poinschw{t_p = -{r \ell e^{-t/\ell}\over (r^2 - \ell^2)^{1/2}}, \qquad r_p = (r^2 - \ell^2)^{1/2} e^{t/\ell}}
The metric becomes
\eqn\hyperBH{ds^2 = -f(r) dt^2 + {dr^2\over f(r)} + r^2 d\sigma^2}
where
\eqn\deff{f(r) = {r^2\over \ell^2} - 1.}\
We can orbifold $\QH_3$ by a discrete group $\Gamma$, leading
to a compact hyperbolic space $\Sigma = \QH_3/\Gamma$.
$d\sigma^2$ now denotes the metric on $\Sigma$.
The metric \hyperBH\ takes the standard form of a static black hole with horizon at $r=\ell$ having constant negative curvature. The inverse transformation, taking \hyperBH\ into \poincmet, is
\eqn\schwpoin{r = -{r_p t_p\over \ell}, \qquad t = - {\ell\over 2}\ln\left({t_p^2\over \ell^2} - {\ell^2\over r_p^2}\right)}
These transformations are valid outside the horizon, i.e., $r > \ell$. Analogous formula apply for $r< \ell$. However, a key point for our later discussion is that the Poincare coordinates $t_p,r_p$ smoothly cover both the region inside and outside the black hole horizon.

The black hole \hyperBH\ has Hawking temperature $T_H = 1/(2\pi \ell)$, while the same spacetime viewed in Poincare coordinates has zero temperature. This is analogous to the statement that Rindler space has nonzero temperature and can be understood as follows.  The dual field theory will be discussed in detail in the following sections, but for now it suffices to note that the metric at infinity in the black hole spacetime is conformal to the static cylinder \staticcyl.
 %
%\eqn\staticcyl{ ds^2 = -dt^2 + \ell^2 d\sigma}
%
 Starting with  Euclidean space
$ ds^2 = dx^2 + dy^2 + dz^2 + z^2 d\theta^2$ and rescaling by $\ell^2/z^{2}$  yields
$ds^2 = \ell^2[d\theta^2 +d\sigma^2]$ which is the static cylinder at temperature $T_H = 1/(2\pi \ell)$. So for any conformal field theory, the standard  Minkowski vacuum is equivalent to a thermal state on this static cylinder\foot{We thank  J. Maldacena for suggesting this argument.}.

As mentioned in the introduction,  a curve of constant Poincar\'e radius $r_p = R$ corresponds to an infalling trajectory in the black hole: $r^2(t) = R^2 e^{-2t/\ell} + \ell^2$. This allows us to construct a simple model of matter collapsing to form the black hole. Consider a static spherical shell of D3-branes.  The metric inside the shell is ten dimensional flat spacetime, while the metric outside is $AdS_5 \times S^5$. If we again introduce Milne coordinates and compactify the hyperbola, this metric becomes
\eqn\shellmetric{ds^2 = h^{-1}[-dt_p^2 + t_p^2 d\sigma^2]  +h[dr_p^2 + r_p^2 d\Omega_5]}
where
\eqn\defh{h(r_p) = {\ell^2\over r_p^2}\ \   (r_p > R), \qquad h(r_p) = {\ell^2\over R^2} \ \  (r_p < R)}
where $R$ is the radius of the shell in Poincare coordinates.  In terms of the black hole interpretation of the exterior spacetime, the shell starts at large $r$ and collapses to $r=0$ forming a horizon at $r=\ell$.

Note that the singularity in \poincmet\  at $t_p=0$ that appears after compactifying the hyperbola is naturally viewed as a cosmological singularity in AdS since it is spacelike and goes out to infinity in finite time. Since this metric is equivalent to  \hyperBH, we are led to the surprising conclusion that there is no invariant distinction between black hole singularities and cosmological singularities in this case. Although this singularity is not a curvature singularity, the slightest perturbation will cause the curvature to diverge.

There is a one parameter family of hyperbolic black holes which generalize \hyperBH. The metric again takes the form \hyperBH\ where now
\eqn\deff{f(r) = {r^2\over \ell^2} - 1 - {\mu\over r^2} }
These are all solutions to Einstein's equation with negative cosmological constant.\foot{For a more detailed discussion of these solutions, see \EmparanGF.}
 These spacetimes describe static black holes with an event horizon at the largest zero of $f$ and a singularity at $r=0$. The constant $\mu$ is a free parameter which is related to the mass of the black hole by\foot{If one computes this mass using the boundary stress tensor this mass is shifted by a ($\mu$ independent) constant  \EmparanGF. Since we are interested in dynamical questions, the zero point of the energy is not important and we will ignore it.}
 \eqn\mass{M = {3\mu \hV \over 16 \pi G_5} }
where $\hV$ is the dimensionless volume of $\Sigma$.
 These solutions have the unusual property that there is a horizon even when $\mu$ is negative provided $\mu\ge \mu_{ext}$ with
\eqn\defextmu{  \mu_{ext} \equiv - {\ell^2\over 4}  \quad \Rightarrow \quad M_{ext} = -{3 \hV\ell^2 \over 64 \pi G_5}}
 For $\mu>0$, the singularity at $r=0$ is spacelike (like Schwarzschild-AdS), but for $\mu_{ext}\le \mu <0$, it is timelike (like Reissner-Nordstrom-AdS). In terms of the horizon radius, $r_+$, the Hawking temperature of these black holes is
\eqn\BHtemp{ T_H = {2r_+^2 - \ell^2\over 2\pi \ell^2 r_+}}
In the extremal limit when $\mu = \mu_{ext}$, the horizon radius is $r_+ = \ell/\sqrt2$ and $T_H = 0$.  For $\mu < \mu_{ext}$ there is a naked singularity.

We believe that one can form these black holes with nonzero mass by throwing in spherical shells of D3-branes with various energies. However, it is difficult to find exact ten dimensional supergravity solutions in this case since the shell will radiate. This does not occur when $\mu=0$ since the shell is following the orbit of a Killing field (Poincare time translations). There is no danger of violating cosmic censorship by throwing in a shell and forming a naked singularity, since we will see in the next section that if  the energy is  sufficiently negative, the shell will bounce, and never reach $r=0$.

We have seen that the $\mu=0$ black hole can be described in Schwarzschild coordinates or Poincare-like coordinates. For our later analysis,  it will be very useful to work with a hybrid system of coordinates which keep the nice time slices of the Poincare coordinates, but for which the boundary theory naturally lives on the static cylinder.  These are given by
\eqn\tildecoords{\tilde r = -{r_p t_p\over\ell} ~~~~~ \tilde t = -\ell\ {\rm ln} \left(- {t_p\over\ell}\right)}
(Note that $\tilde r$ is the same as the Schwarzschild radial coordinate $r$.)  In these coordinates, the metric for the $\mu=0$ black hole is
\eqn\tildemetric{ds^2=-\left({\tilde r^2\over\ell^2}-1\right)d\tilde t^2+\tilde r^2d\sigma^2+{\ell^2\over\tilde r^2}d\tilde r^2+{2\ell\over\tilde r}d\tilde t d\tilde r}
These coordinates extend smoothly across the horizon.

\newsec{The Dual CFT}

\subsec{CFTs on cosmological vs. static backgrounds}

In order to study the systems discussed in the previous section non-perturbatively, we will
analyze the dual CFT.  This CFT lives on a non-fluctuating background spacetime which is homogeneous along hyperbolic spatial slices.  Since the ${\cal N}=4$ SYM theory is conformally invariant, we should obtain the same results from the field theory in any conformal frame, modulo subtleties arising in the case of singular conformal transformations.  There are two conformal frames which will prove useful to consider, as discussed in \GreenTR.
First, consider the CFT on the spacetime \ccgeom\
\eqn\FRWframe{ds^2=-dt_p^2+t_p^2d\sigma^2}
with $d\sigma^2$ the metric on $\QH_3$. This is simply a hyperbolic slicing of flat spacetime, and  has vanishing curvature.

In the presence of additional stress-energy which is homogeneously distributed along the hyperbolic slices, however, the system becomes singular as $t_p\to 0$. Upon orbifolding by $\Gamma$,
so that $d\sigma^2$ is the metric on $\Sigma$, then even in the absence of additional sources of stress energy, the spacetime \FRWframe\ has a Milne-like singularity at  $t_p=0$.  The dual bulk geometry is the region of spacetime covered by the metric \poincmet.

Next, consider the cooordinate transformation of \FRWframe\ (with $t_p < 0 $) to conformal time $\eta$:
\eqn\conftransf{t_p=-\ell\ e^{-\eta/\ell}
%~~~~~ r = -t_p r_p ~~~~
}
which yields $ds^2 = e^{-2\eta/\ell}(-d\eta^2+\ell^2 d\sigma^2)$. We now conformally rescale the metric:
\eqn\staticframe{ds^2_{static}= e^{2 \eta/\ell} ds^2 = -d\eta^2+\ell^2 d\sigma^2}
and the operators of the CFT.  This maps the CFT on \FRWframe\ to that on \staticframe. This transformation is singular at $t_p=0$. The static metric $ds^2_{static}$  has constant negative scalar curvature ${\cal R}$ coming from the hyperbolic spatial slices.
This static background spacetime has no singularity whether or not we compactify the space or include homogeneous sources.  We will argue below that the singularity in the spacetime physics will manifest itself in this static background through the behavior of the rolling scalar fields of the CFT.

\subsec{Yang-Mills theory and scalar dynamics}

The ${\cal N}=4$ SYM theory has six adjoint scalar fields $\Phi^i$, which together with the gauge fields are governed by the classical action
\eqn\CFTac{{\cal S}_{SYM}=\frac{-1}{g_{YM}^2}\int \sqrt{-g}Tr(F^2+(D\Phi)^2+ [\Phi^i,\Phi^j]^2+
\frac{1}{6} {\cal R}\Phi^2)+fermions}
These scalars, and their gravity-side manifestation as collective coordinates for D3-branes, will play a key role in our analysis.
The commutator interaction among the scalar fields is minimized by commuting matrices, i.e. eigenvalues $\phi^i_a,a=1,\dots N$ of the $\Phi^i$.  In  the case of flat spacetime, this leads to a moduli space of scalar field VEVs, homogeneous along flat spatial slices; otherwise generically the moduli space is only approximate.  The off-diagonal modes of the $U(N)$ adjoint matrices (the ``W bosons") have masses
\eqn\Wmass{m^2_{W,i ab} \sim g_{YM}\sum_{j\ne i}(\phi^j_a-\phi^j_b)^2}
They all become light at the origin $\phi^i_a\to 0$.

%On more general spacetime backgrounds, such as those with homogeneity along hyperbolic spatial slices, the space of homogeneous scalar fields is not generically a flat moduli space.  The nontrivial background geometry affects the normalizability and the effective action for the various modes of the scalar fields.  On noncompact hyperbolic space, the zero mode of a free scalar field is non-normalizable.  For compact hyperbolic space, the zero mode is normalizable, and since  in \CFTac\ the scalar fields are conformally coupled, an eigenvalue $\phi_a$ is subject to a quadratic potential ${\cal R}\phi_a^2$ far out in field space.

Consider first this quantum field theory on \FRWframe.  This is a CFT on a Friedmann-Robertson-Walker background or ``collapsing cone" with scale factor $a(t_p)=t_p$.
For large $|t_p|$, $H=\frac{\dot a}{a}\sim 1/t_p$ is small and the physics is smooth.  For small $|t_p|$, there is a singularity; Kaluza-Klein modes on the hyperboloid blueshift without bound and the ratio $m_W/m_{KK}\propto |t_p|$ goes to zero.

Next, let us analyze this system in the static frame \staticframe.  The conformal transformation acts on the scalar fields as
\eqn\scalartransf{\Phi = e^{-\eta/\ell} \Phi_p}
where $\Phi$ refers here to the field in the static conformal frame.  Applying this to a diagonal element of $\Phi$, we see from  \Wmass\ that in this frame $m_W\to 0$, as the scalar field reaches the origin,
while the KK mass is constant.   Thus, both frames contain a regime of the dynamics in which
the dimensionless ratio $m_W/m_{KK}\propto |t_p|$ goes to zero in this limit, and these
regimes should be identified.

In the static frame, the conformal coupling of the scalar field to the scalar curvature ${\cal R}$ gives a negative mass squared to the scalars. The theory does not have a ground state.\foot{Of course, the
theory on the collapsing cone does not have a unique vacuum either.} Nonetheless, the gauge theory is well defined.  First, only a finite number of modes (and in some cases, only the zero modes \CornishHZ) of the scalars experience this instability.
Secondly, it takes an infinite time for these zero modes to roll down a $- Tr \Phi^2$ potential to infinity. The classical and quantum evolution is well defined for any initial state. Finally, the physics of the scalars rolling off to infinity is equivalent to the scalars being
constant at late time on the {\it expanding}\ cone, a regime which is completely  well behaved.

We deduce that in the static frame \staticframe, the physics of the singularity of the metric \FRWframe\ corresponds to scalar fields rolling toward the origin $\Phi=0$ and scattering off a quadratic potential barrier.  Since both the collapsing cone metric and the conformal transformation to the static frame that we used are singular, we will focus on the field theory formulated in the static frame, and study the relationship between the spacetime and the field theory pictures.

To begin with, let us consider gauge theory states
at different energies $E$, and map them to the corresponding family of backgrounds on the gravity side.
For energies $E > 0$, one can argue that the generic state of the field theory will be
an excited gas of W-bosons and Kaluza-Klein modes, with the scalars trapped for
a long time at the origin.  The gravitational duals will be the $M > 0$ black holes.
Negative energies corresponding to $M < 0$ can be accessed by lowering the
energy of the zero modes of the scalar fields; the potential barrier due to the curvature
energy means that the theory will be out on the Coulomb branch.  More generally,
as noted in \S2, we can create black holes at all energies, by studying shells of N D3-branes
wrapped along $\Sigma$, and distributed evenly over the $S^5$.
Out on the Coulomb branch, the eigenvalues of $\Phi$ are naturally associated
with the radial positions of the dual D-branes.

We find a simple correspondence between the quantum corrections in the  scalar field theory and the
causal structure in the gravitational duals.
%Classically an eigenvalue $\phi$ rolling toward the origin moves at a constant velocity $\dot\phi$.
In the static frame, the eigenvalues for the scalar fields see an inverted harmonic oscillator potential.
If the off-diagonal modes are set to zero (so that the commutator term in \CFTac\ does not contribute), then there are classical trajectories in which each eigenvalue $\phi$ of energy $E_\phi\sim -\phi_0^2\ell/g_{YM}^2$ executes a motion
\eqn\classtraj{\phi(t)=\phi_0 \cosh[(\eta- \eta_0)/\ell]  }
with $\phi$ bouncing off the inverted harmonic oscillator potential at a minimal value $\phi=\phi_0$
when $\eta = \eta_0$.
With $N$ eigenvalues moving in an $SO(6)$ invariant configuration, this has energy
\eqn\Eclasstraj{E\sim -{N\phi_0^2\ell\over g_{YM}^2}}

Near the origin, the off-diagonal modes of the $U(N)$ matrices become light, and generate
quantum mechanical effects which drastically affect the evolution.
At weak coupling ($\lambda\ll 1$),
particle production of off-diagonal modes (``W bosons") traps the field at $\phi=0$
\refs{\DouglasYP,\KofmanYC}. The strength of this effect is controlled by
\eqn\partprod{{1\over{\phi^2}}|{d\phi\over{ d\eta}}| \sim {1\over\phi_{0}\ell}}
At strong 'tHooft coupling, the leading effect is the renormalization of the action
due to loops of W bosons, which sum up to the DBI action and slow and trap the field at $\phi=0$ \SilversteinHF.
The figure of merit for the importance of loop effects arising from the time dependence in the rolling scalar background\foot{In the case of present interest, the field theory is not formulated on Minkowski space and the DBI action in general also contains corrections arising from the curvature of the background metric and from finite temperature effects.} is $\sqrt{\lambda}$ times \partprod:
\eqn\DBIcorr{{\sqrt{\lambda}\over{\phi^2}}|{d\phi\over{ d\eta}}| \sim {\sqrt{\lambda}\over\phi_{0}\ell}}

For sufficiently negative energy $E$ the field bounces off the inverted harmonic oscillator potential at a field value $\phi_0$ sufficiently large that these effects never become important.  The loop
effects become important when the minimal field value is small enough that \DBIcorr\ becomes $\ge 1$, i.e. when $\phi_0\ell \le \sqrt{\lambda}$.  Translating this into an energy scale using \Eclasstraj, this implies $E>-N^2/\ell$.  This energy scale is significant on the gravity side:  from \defextmu\ and the
fact that $\frac{\ell^2}{G_5} \sim \frac{N^2}{\ell}$ we see that it is parametrically the energy scale above which there is a black hole horizon
%and singularity
in the dual geometry.  Therefore the field theory produces nontrivial DBI corrections precisely in the regime where the classical gravity side exhibits nontrivial causal structure.\foot{In \refs{\KabatVC}, the
authors have argued that for Schwarzschild observers, the breakdown of the DBI action of D-brane probes at the horizon coincides with the appearance of light states connecting the
probe and the hot Dp-branes.  It would be interesting to see if a similar picture here led
to thermalization of a collapsing shell as it approached the horizon.}

\ifig\phasediag{Phases of the scalar field dynamics. The curve is the
inverted harmonic oscillator potential, the straight lines are the incoming scalar field
trajectories at various energies, and the dashed lines are the classical late time
behavior (meant to be at the same energy) without quantum corrections.
Trajectory (A) corresponds
to a bouncing scalar.  In trajectory (B), the scalar would bounce without quantum corrections,
but DBI corrections slow the scalar near the potential barrier and quantum corrections become
important. In trajectory (C), the scalar would sail over the potential barrier without quantum corrections,
but again DBI and other quantum corrections become important near the origin.
} {\epsfxsize3.0in\epsfbox{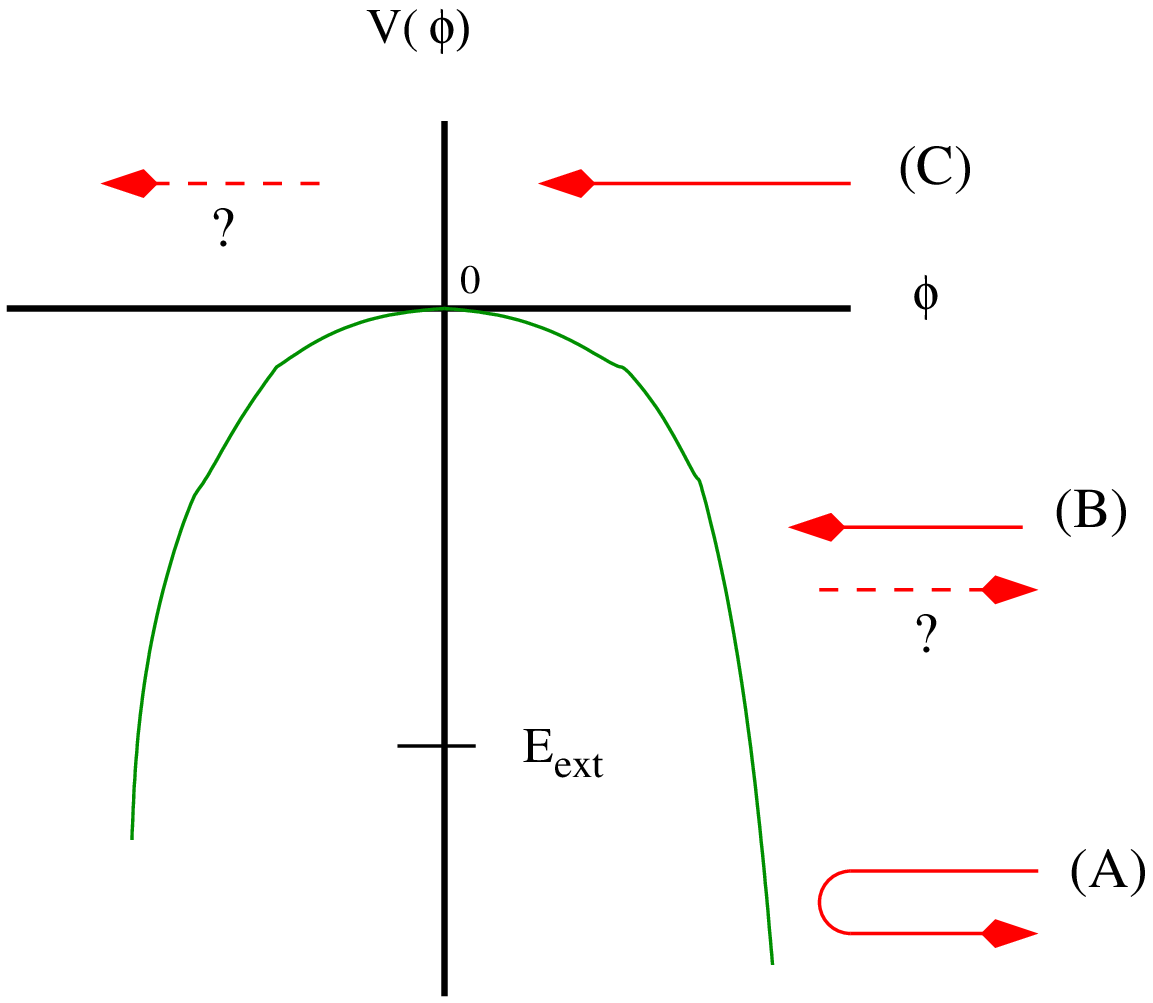}}

In general, the dynamics of the scalar fields have a phase structure which matches the different
causal structures for the dual spacetimes, as seen in Figures 2 and 3.
Let us collect the above statements to make this clear. For $E < - N^2/\ell$, the scalar
fields bounce off of the potential before
quantum effects are important.  The dual spacetime has no horizons, and we believe that the naked singularity is removed by the D-brane shell along the lines of \refs{\JohnsonQT}. For
$- N^2/\ell < E < 0$, the scalar field trajectories see strong quantum corrections before
they reach the potential barrier, roughly at the location of the horizon; furthermore, the barrier prevents them from reaching the origin. The dual spacetimes are black holes of Reissner-Nordstrom type,
with inner and outer horizons cloaking timelike singularities (although the inner
horizon is unstable to forming a singularity). Finally, for $E > 0$, the scalar field trajectory continues
to be corrected when the scalars reach the horizon; furthermore, there is no barrier
preventing them from reaching the origin.  In this regime, the dual spacetimes are
black holes with {\it spacelike}\ singularities.

\ifig\causalstructure{Causal structures of classical black holes formed by scalar trajectories shown in the
previous figure.  For case (A), the brane bounces and the putative naked singularity in
\hyperBH\ for $E < E_{ext}$ is screened.  For (B) we are unsure of the trajectory of
the collapsing shell, and have shown the Reissner-Nordstrom-like
causal structure of the black holes \hyperBH\
for $0 > E > E_{ext}$.  For (C) the black holes have spacelike singularities, and we have shown
the diagram for a $M=0$ black hole.}{\epsfxsize4.0in\epsfbox{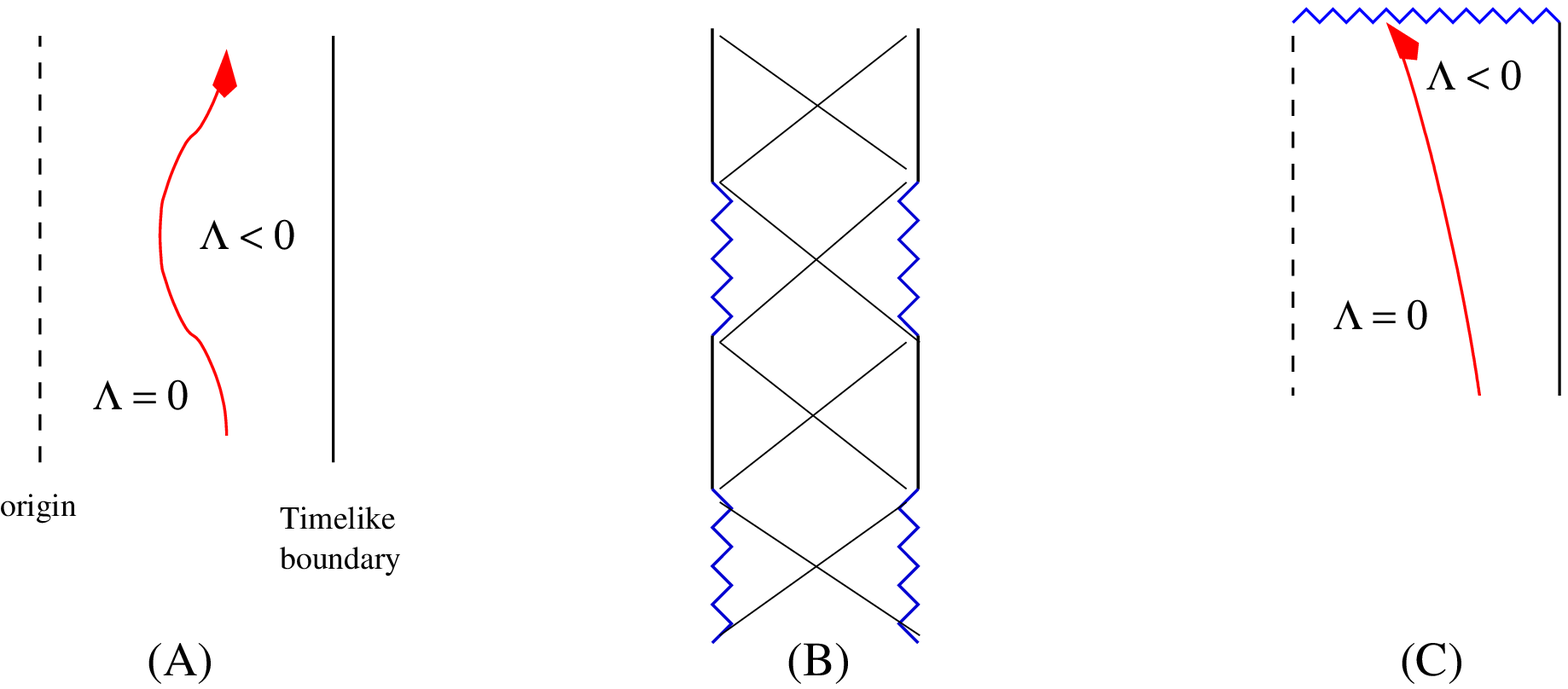}}
%

%in Schwarzschild coordinates
%One argument in favor of this identification will become clearer in \S4; the leading terms
%in the DBI action for the D0-branes at large $r$ match the leading terms in the Yang-Mills action \CFTac.
%For $M=0$ black holes, the identification of $r_p$ with the scalar field values for Yang-Mills on
%\ccgeom\ is inherited from the known identification for Poincar\'e coordinates.  Since the action of
%\scalartransf\ on $r$ is identical to \schwpoin, $\phi$ and $r$ should be identified for the theory on
%\staticcyl\ as well.\foot{It is likely that at finite temperature, a finite field redefinition will
%be required to match $r$ and $\phi$ near the horizon, as in \refs{\MaldacenaKR}.}

%For any energy $E \equiv M$, the gravity solution for $r > r(t)$ will be described to good approximation
%by the metric \hyperBH, with $f$ given by \deff\ and $M$ given by \mass.
%\foot{More precisely, there will be some gravitational radiation as well on this background.} Inside
%the shell, the RR flux will be screened and the cosmological constant will vanish.

%However, before turning to that, let us first explain
%some basic features of the physics of the singularity (=origin of CFT moduli space).

The stalling of the system at small values of $\phi$ appears to be key to understanding the resolution of the singularity and the description of the decay of the black holes \hyperBH.
There are many coupled features of the physics in this regime, which we will only partially control.  These include the loop effects discussed above, production of KK modes and W bosons, the Casimir energy (which in general breaks hyperbolic symmetry \refs{\MullerPE}), the dynamics of Wilson lines (both continuous and discrete), and the spreading of the wave function for the scalars.

In the present work, we will largely focus on the CFT description of the process of an observer propagating through the horizon.  The first step at large $\lambda$ is to take into account the effects of the DBI action.  This will lead us to a scheme for describing physics inside the horizon, accessible before reaching the full complications of the singularity. In particular, in order to analyze the dynamics of the scalar fields at strong coupling, we can make use of the gravity side of the AdS/CFT correspondence in the regime where it is effective (i.e. away from the singularity). Let us turn now to these issues. In  \S5, we will explore the physics of the singularity and the fate of the black holes in light of its relation to the dynamics of the Yang-Mills theory at the origin of moduli space.

%To work in the first frame \FRWframe\ above, we consider a D3-brane at a radial position $r_p$ in the Poincare patch %of $AdS_5$, which is foliated by flat four-dimensional slices \FRWframe.  Because of the vanishing of the curvature of %\FRWframe, the conformal coupling ${\cal R}\phi^2$ vanishes in this frame.
%The effective action for a given eigenvalue of the $U(N)$ matrix of scalar fields is the DBI action (in the absence of %significant acceleration in the bulk).  This action is
%
%\eqn\DBIPoinc{S_{DBIP} = -\int d^4 x {r_p^4 t_p^3\over{\ell^4 g_s\alpha'^2}} \left(\sqrt{1-\frac{\ell^4 (\dot %r_p^2-{\ell^2\over t^2}(\nabla r_p)^2)}{r_p^4}}-1\right) }
%
%Varying this action with respect to the radial collective coordinate $r_p$, one obtains an equation of motion
%
%\eqn\DBIPoinceom{4 r_p ({1\over\gamma}-1)+{3\dot r_p\gamma\over t_p}+{d\over{dt_p}}(\dot r_p\gamma)-2{\dot r_p^2\over %r_p}\gamma = 0}
%
%where $\gamma=1/\sqrt{1-\ell^4\dot r_p^2/r_p^4}$.
%This has a constant-$r_p$ solution which describes a D3-brane falling into the black hole (figure).  The second term %here contains the effects of Hubble friction on the motion, following from the homogeneity along the hyperbolic as %opposed to flat spatial slices of the Poincar\'e slices of $AdS_5$.

%Making a conformal transformation
%
%\eqn\conftransfP{t_p=-e^{-\tilde t}  ~~~~~~ r_p\equiv\rho e^\eta}
%

\newsec{How to look behind the horizon}

For gauge theories with a gravitational dual, the theory on a static 4d spacetime at finite temperature is taken to describe the experience of a ``Schwarzschild" observer
in the dual 5d static black hole spacetime. Classically, such an observer cannot see behind the horizon.
When the black hole is formed dynamically, unitarity requires that the experience of infalling observers
behind the horizon is measured by a Schwarzschild observer through subtle correlations in the black
hole's Hawking radiation.\foot{Different language describes the
consequences of unitarity for ``eternal" black holes, whose field theory dual is  a tensor product
of two copies of the zero-temperature field theory\ \refs{\HorowitzXK,\BalasubramanianDE,\CarneirodaCunhaJF,\MaldacenaKR}, coupled via a correlated ``thermal" state.}

The sections above give a prescription
for describing observers such as D3-brane probes falling through the
horizon of $M=0$ hyperbolic black holes:
study the gauge theory
%in the "collapsing cone" background \FRWframe.
dual to the $t_p<0$ half of the Poincar\'e patch.
How can we relate this to the experience of a Schwarzschild observer?
%, whose experience
%is well-described by the gauge theory on the 4d spacetime \staticframe?
In this section, we give an answer for infalling D3-brane probes wrapping
the hyperbolic space $\Sigma$.
The essential point is that one may promote the time direction to a field
variable of the D-brane action, which amounts to adding  an additional
gauge invariance (invariance under worldvolume reparametrizations).
The resulting DBI effective action for the D3-brane probe
is invariant under {\it target space} reparametrizations.  One can
transform the theory from a description suited to different observers by  simple
field redefinition. This answer generalizes readily to $M\neq 0$ black holes, and
to black holes with other horizon topologies, such as those with spherical or flat horizons which
appear in asymptotically AdS backgrounds.

We will start in \S4.1\ by describing the the DBI actions for
D3-brane probes according to both ``infalling" and ``Schwarzschild" observers, and constructing
a probe action for infalling observers which is dual to a non-singular gauge theory.
In \S4.2\ we will discuss the change of variables of the probe theory which transforms
these actions into each other. These maps take the form of a field-dependent
reparametrization of the gauge theory time variable. In \S4.3\ we will argue that
these reparametrizations may arise from a change of gauge in the underlying Yang-Mills theory.

\subsec{The effective action for D3-brane probes}

Consider a single D3-brane wrapping $\Sigma = \QH_3/\Gamma$, with
the worldvolume time set equal to the Poincar\'e time and the position described by
the radial direction in Poincar\'e coordinates.  The DBI action for this
D3-brane probe is:
\eqn\dbip{
	S_{dbip} =  \frac{\hV}{g_s (\alpha')^2} \int d\tp \left[\frac{\rp^3 \tp^3}{\ell^3}
		\sqrt{\frac{\rp^2}{\ell^2} - \frac{\ell^2}{\rp^2} {\dot \rp}^2} - \frac{\rp^4\tp^3}{\ell^4}\right]\ .
}
(Here $\hV$ is the volume of the compact hyperbolic space).
Gauge-gravity duality tells us that \dbip\ arises from the
Yang-Mills theory on the cosmological spacetime
\FRWframe, with one eigenvalue $\phi_p = m_s^2 r_p$ of the adjoint scalar taken far out along the Coulomb branch, breaking $U(N) \to U(N-1)\times U(1)$; it is the effective action
for this eigenvalue, which arises from integrating out the massive W bosons with mass
$m_W = \phi_p$. In particular, one might have deduced this action by
considering the AdS/CFT correspondence in Poincar\'e coordinates, and performing
the orbifold on that theory, truncating to zero modes after the coordinates
have been changed and the orbifold projection has been performed.

This action is well suited to describing an infalling observer (where the observer is
our D3-brane probe). The equation of motion for $r_p$ is:
\eqn\DBIPoinceom{4 \frac{r_p^3}{\ell^4} ({1\over\gamma_p}-1)+{3\dot r_p\gamma_p\over t_p}+{d\over{dt_p}}(\dot r_p\gamma_p)+2{\dot r_p^2\over r_p}\gamma_p = 0}
where $\gamma_p=1/\sqrt{1-\ell^4\dot r_p^2/r_p^4}$.
This has a constant-$r_p$ solution which describes a D3-brane falling into the black hole, reaching the singularity at $t_p=0$, which is at finite time.  The second term here contains the effects of Hubble friction on the motion, following from the homogeneity along the hyperbolic as opposed to flat spatial slices of the Poincar\'e slices of $AdS_5$. While DBI corrections may begin
to become important near the horizon, it is easy to check that fluctuations around this solution
remain under control all the way through the horizon.  As we will discuss at the end of this section, it is useful to describe this in a static conformal frame obtained by the conformal transformation \staticframe\ to the static cylinder; this leads to a description of infalling observers related to the coordinates $\tilde r,\tilde t$ \tildecoords.

On the other hand, we also wish to describe an observer who remains outside of the black hole,
and so study the detailed implementation of black hole complementarity from the field theory
point of view. Furthermore, as stated in the introduction of this section,
past presentations of black holes are in terms of a gauge theory on a static background at finite temperature, which describes Schwarzschild observers, and we would like to study physics
behind the horizons of these black holes as well.

Consider then a D3-brane wrapping $\Sigma$, with worldvolume time equated with
Schwarzschild time, and the position described by the radial position in Schwarzschild
coordinates. The DBI action describing the spacetime dynamics of this probe is\foot{The last term comes from $\int A_4$ where $A_4$ is the potential for the five-form flux. In Schwarzschild coordinates, the potential which is nonsingular on the horizon is $A_4 \propto (r^4 - \ell^4) dt\wedge \epsilon_3$ (where $\epsilon_3$ is the volume form on the hyperboloid). Starting with the form of  $A_4$ in Poincare coordinates and applying the coordinate transformation
\poinschw\ yields an expression for $A_4$ which agrees with this up to a term $dA_3$ where $A_3$ is finite on the horizon.}
\eqn\dbis{
	S_{dbis} = - \frac{\hV}{g_s (\alpha')^2} \int dt \left[ r^3 \sqrt{f(r) - \frac{{\dot r}^2}{f}} - \frac{(r^4 - \ell^4)}{\ell}
	\right]\ ,
}
Gauge-gravity duality tells us
that this action arises from considering $\CN=4$ $U(N)$ Yang-Mills theory
on \staticframe\ at finite temperature, with one eigenvalue $\phi = m_s^2 r$ of the adjoint scalar
taken out onto the Coulomb branch, again breaking the $U(N)$ gauge symmetry to
$U(N-1)\times U(1)$.  $S_{dbis}$ is the effective action for this eigenvalue, arising
from integrating out the W-bosons (transforming in the fundamental of the unbroken $U(N-1)$)
with mass $m_W = \phi$.

Near the horizon, the scalar field approaches its speed limit
\eqn\speedlimitS{\dot r^2\to f(r)^2,}
and the relativistic $\gamma$ factor $\gamma=1/\sqrt{1-\dot r^2/f^2}$ approaches infinity.
Therefore, the probe takes an infinite Schwarzschild time to reach the horizon, as is
clear from the time slicings depicted in Figure 1b. Furthermore,
perturbations about the background solution become large near the horizon, for the following reason.  Expand $r\equiv r_0(t)+\delta r$ about a spatially homogeneous classical solution $r_0(t)$.  The resulting effective action for the perturbations $\delta r$, obtained by expanding the square-root action, has a quadratic term scaling like $\gamma^3$, cubic interactions scaling like $\gamma^5$, and so on.  Therefore the interactions among the perturbations $\delta r$ become strong as the brane approaches the horizon.
%corrections
%to the quadratic action diverge (due to the $1/f$ term multiplying $\dot{r}^2$),
%and quantum fluctuations around $r(t)$ become large.
This suggests that \dbis\ is not a good effective description near the horizon, and
one should choose a different set of variables to describe the probe dynamics.

Such variables are provided by our original Poincar\'e description.
The reader may worry that as we originally formulated it, the gauge theory dual which describes infalling
observers well lives on a singular spacetime.  In fact, we
can choose field theory variables for the gauge theory on the static cylinder which
sees behind the horizon, giving us the best of both worlds. Furthermore,
this set of fields will be most readily generalized to other black holes.

These variables are just the tilded coordinates introduced at the end of \S2. The DBI action in the metric \tildemetric\ is
\eqn\Poincaretilde{S_{dbip}'=- \frac{\hV}{g_s (\alpha')^2} \int d\tilde t{\tilde r^4\over {\ell}}
\left\{\sqrt{1-{\ell^4\over \tir^4}\left ({d \tilde r\over {d\tilde t}}+{\tilde r\over \ell} \right )^2}-1\right\}}
Since \conftransf,\staticframe\ map the collapsing cone to the static cylinder, this theory
should also arise from Yang-Mills on the non-singular static cylinder.

In these coordinates, the late-time solutions approach the speed limit
\eqn\speedlimitPoinctilde{\left ({d \tilde r\over {d\tilde t}}+{\tilde r\over \ell} \right )^2\to {\tilde r^4\over\ell^4} }
In other words, all solutions asymptote to $\tilde r\to \tilde r_0 e^{-\tilde t/\ell}$ at late times $\tilde t$.
In these variables, the classical approximation breaks down at the singularity: as in \dbip, the evolution through the horizon is smooth and classical.  This should
not surprise us: the transformation \tildecoords\ reparametrizes the
equal-time slices in Poincar\'e coordinates slice by slice, but does not change the slicing.
The upshot is that we can separate the problem of the spacetime singularity from the
problem of studying Yang-Mills theory on a nonstatic, singular spacetime.

The reader may also object that we have argued in \S3\ that the physics of the singularity
is captured by the origin of moduli space, which the Schwarzschild coordinate $r(t)$ cannot classically reach.
Black hole complementarity states that the Schwarzschild observer should somehow be able
to access the physics near the singularity, if only through subtle correlations in the radiation
emitted by the black hole.  Moreover, it would be worrisome if a regime of field space was somehow excised.  However, we believe that while the classical D3-brane probe
reaches the horizon only as $t\to\infty$, these probes and the field theory degrees of
freedom which couple to them become strongly fluctuating in Schwarzchild variables.
Such fluctuations may extend to the origin of moduli space;
but they will have no classical spacetime interpretation. A Schwarzschild
observer will see infalling objects approach the horizon and thermalize.\foot{It
is possible that a story analogous to \KabatVC\ pertains here.}

In \S4.2, we will describe the transformation
of the quantum variables of the low energy effective action, which maps \dbip\ to \dbis.
However, the full Yang-Mills theory on the static cylinder
seems to produce two effective actions. On the one hand, the Schwarzschild
DBI action \dbis\ seems natural for a theory on the static cylinder at finite temperature, by analogy
with other black holes. On the other hand, since \conftransf\staticframe\ map the Yang-Mills actions on
\FRWframe,\staticframe\ into each other, it would appear that \Poincaretilde\ should arise naturally
from the Yang-Mills theory.  This is not a contradiction: the definitions of time in \dbis,\Poincaretilde\
differ for finite field values.  Furthermore, the form of the effective action will depend on
the coordinates used on the space of eigenvalues, and on the choice of gauge used
(since the eigenvalues are not gauge-invariant objects). In \S4.3\ we will argue that that the difference
between \dbis\ and \Poincaretilde\ arises at least in part from the action of
\conftransf\staticframe\ on the gauge condition used to compute the DBI action.

\subsec{Time reparameterizations in quantum mechanics}

The basic idea behind the transformation of quantum variables describing the Schwarzschild and
infalling observers is simple.  Beginning with either \dbip, \dbis, or \Poincaretilde, one may
promote the corresponding times $t_p, t$ to quantum variables $t_p(\tau)$, $t(\tau)$ at the
price of introducing a gauge invariance -- worldvolume reparametrizations --
which must be dealt with in the path integral.  The resulting DBI action of a particle is written as:
\eqn\dbiact{
	S \propto \int d\tau \sqrt{- \det G_{\mu\nu} \p_{\tau} X^{\mu} \p_{\tau} X^{\nu}} + S_{WZ}\ ,
}
where $S_{WZ}$ is the ``Wess-Zumino" coupling to background Ramond-Ramond fields.
In addition to the aforementioned gauge invariance, this action is also manifestly invariant under
target space coordinate transformations.\foot{Up to boundary terms, which can be removed by gauge transformations of the Ramond-Ramond fields.} In particular, \dbis\ becomes:
\eqn\dbisc{
	S_{dbisc} = - \frac{\hV}{g_s \ell_s^4} \int d\tau \left[ r^3 \sqrt{ f t_{,\tau}^2 - \frac{r_{,\tau}^2}{f}}
		- \frac{(r^4-\ell^4) t_{,\tau}}{\ell} \right]
}
and \Poincaretilde\ becomes:
\eqn\dbiscprime{
	S_{dbisc'} = - \frac{\hV}{g_s \ell_s^4} \int d\tau \left[ \tir^3
	\sqrt{\frac{\tir^2}{\ell^2}\tdt_{,\tau}^2 - \frac{\ell^2}{\tir^2}
		\left(\tir_{,\tau} + \frac{\tir}{\ell}\tdt_{,\tau}\right)^2} - \frac{\tir^4\tdt_{\tau}}{\ell}\right]
}

The transformation between \dbisc\ and \dbiscprime\ is clear. Since \dbiact\ is invariant under
target space coordinate transformations, we need merely find the coordinate transformation
that relates the spacetime coordinates $r,t$ to $\tir, \tdt$, and promote these
to field redefinitions.  These transformations are:
\eqn\sprimetos{
	\tir = r \ ; \ \ \ \ \tdt = t + \frac{\ell}{2}\ln\left(1 - \frac{\ell^2}{r^2}\right)\ .
}

In the covariant DBI action \dbiact, this is just a (complicated)
field redefinition. After transforming the action, one may fix gauge again by selecting a time variable
and setting it equal to $\tau$. In particular, \dbis\ arises by writing \dbiact\ in the
form \dbisc\ and setting $t = \tau$; \Poincaretilde\ arises
by writing \dbiact\ in the form \dbiscprime\ and setting $\tdt = \tau$.\foot{Of course, \dbis\ and \Poincaretilde\ arise from a wrapped D3-brane, so that one must start with a 3+1-dimensional DBI action and fix the gauge for spatial reparametrizations as well; however, since the spatial directions along
$\Sigma$ do not figure into our coordinate transformations, we will ignore this step.}

From the point of view of the gauged-fixed actions \dbis,\Poincaretilde, these transformations
look exotic. $r,\tir$ are quantum observables, so that the second equation in \sprimetos\
appears to be an operator-valued reparametrization of the time variable.  On the other hand,
in theories with reparametrization
invariance, this kind of transformation is common.  For example, when computing
the density fluctuations which arise during inflation, one can choose spatially flat equal-time
slices and study quantum fluctuations of the inflaton; or one may choose a gauge in which the
inflaton fluctuations vanish and a scalar mode of the spatial metric fluctuates. These are related
by a field-dependent change of the equal-time slicing.

Such field redefinitions and gauge transformations leave the physics
invariant.  In practice, they will turn a simple correlation function in one set of variables into a correlation function of very complicated operators in the other set of variables. The field redefinitions and gauge transformations we perform will change which set of operators are
natural to study.  Certain questions will be naturally answered in a frame for which the
corresponding fields are well described by a classical limit. For observers falling into the black hole,
the appropriate frame is realized by the Poincar\'e coordinates and associated
DBI actions \dbip\Poincaretilde, which describe semiclassical evolution of a
probe brane through the horizon.

There has been speculation that to follow an object falling through a black hole horizon,  one should evolve the probe with a new Hamiltonian in the gauge theory. This is indeed what happens in the familiar case of crossing the Poincar\'e horizon by transforming from  Poincare to global coordinates. The Hamiltonian in the gauge theory on $S^3 \times \QR$ is related to the dilatation operator in Minkowski spacetime. However there is a key difference between this and the case we are considering here: The choice of time on the boundary changes when we transform Poincar\'e into global coordinates. In our present example, the choice of time on the static cylinder does not change. The constant $\eta$ surfaces on the boundary can be extended inside the bulk spacetime either along constant $t$ or constant $\tilde t$ time slices.

In the present case, despite the nontrivial change in bulk time slices, the
Hamiltonians in the frames \dbis,\Poincaretilde\
are the {\it same}.  To see this, we can use \dbisc\ and \dbiscprime\ to compute the conjugate momenta.
It is easy to show that:
\eqn\momentamap{
\eqalign{
	p_{\tir} & = p_r - \frac{\ell}{r f(r)} p_t + \frac{r^4 - \ell^4}{\ell^4 r f(r)} \hV N \cr
	p_{\tdt} & = p_{t} + \frac{\hV N}{\ell}\ .
}}
While the momenta conjugate to $r,{\tilde r}$ transform nontrivially into each other,
$p_t$ and $p_{\tdt}$ are equal up to an overall constant shift.  Now upon choosing the gauge
$t = \tau$, $- p_t$ becomes the Hamiltonian derived from \dbis; while choosing the gauge
$\tdt = \tau$, $- p_{\tdt}$ becomes the Hamiltonian derived from \Poincaretilde.
Eq. \momentamap\ shows that these Hamiltonians are the same up to a constant which
is unimportant for the field theory dynamics.

On the other hand, because $p_r$ and $p_{\tir}$ do transform nontrivially, the
functional form of the Hamiltonians will be quite different in different frames.
The reparametrization invariance implies a nontrivial constraint on the canonical
variables.  This can be written as a relation between the conjugate momentum $p_t$
and the variables $p_r, r$. Using the fact that $p_t $ is minus the Hamiltonian (in the gauge $\tau = t$) we find:
\eqn\schwconstr{
	H(p_r,r) = \frac{\hV}{g_s(\alpha')^2}\left[  f(r) \sqrt{\left(\frac{g_s (\alpha')^2 p_r}{\hV}\right)^2 + \frac{r^6}{f}} - \frac{(r^4-\ell^4)}{\ell}
		\right]
}
This can also be obtained by computing the Legendre transformation of \dbis\ to find the Hamiltonian directly.  Similarly, using the fact that
$p_{\tdt}$ is minus the Hamiltonian in the gauge $\tdt = \tau$, we find:
\eqn\tildeconstr{
	H(p_{\tir}, \tir) = \frac{\hV}{g_s (\alpha')^2} \left[  \frac{\tir^2}{\ell^2}
			\sqrt{\left(\frac{g_s (\alpha')^2 p_{\tir}}{\hV}\right)^2 + \ell^2 \tir^4} - {{\tilde r} \over \ell}p_{\tir} - \frac{\tir^4}{\ell}\right]
}
Note that the form of $H$ in terms of the new canonical variables $p_{\tir}, \tir$ is different.
It would be interesting to pursue these transformations further, for example by comparing the
evolution of wavepackets in each set of coordinates.

An important question is how these transformations
lift to the full set of variables of the underlying Yang-Mills theory.  Since the non-abelian
generalization of the DBI action is not known beyond low orders in velocities, this may prove
somewhat difficult.  We now turn to an alternate route to understanding the transformation
of the full Yang-Mills theory.
			
\subsec{Coordinate transformations vs. gauge transformations}

\def\tG{{\tilde G}}

The Yang-Mills actions on \staticframe\ and \FRWframe\ are related by a conformal transformation. The DBI action \dbip\ transforms under this conformal transformation
to \dbiscprime\ (with $\tau = \tdt$), not to \dbis: one requires the additional field-dependent
coordinate transformation \sprimetos\ to pass from \Poincaretilde\ to \dbis.

On the other hand, the effective action \dbis\ naturally captures thermalization in the gauge theory,
and it is clear that any calculation using a gauge, regulator,
and background field configuration which respects the time-reversal symmetry of
the Yang-Mills theory and of the thermal state cannot produce \Poincaretilde\ as an effective action.
For example, the expansion in $(\ell/\tir)^2 \sim \lambda/(\ell\tilde{\phi})^2$
becomes a power series in $(\dot{\tir} + \frac{\tir}{\ell})^2$
rather than $\dot{\tir}^2$.

So either one of \dbis,\dbip\ does not arise from the
Yang-Mills theory on \staticframe,\FRWframe\
respectively, or additional input is required to derive a particular form of the DBI action
as an effective action. An important input is the choice of $U(N)$ gauge. The form of the 1PI
effective action depends on the choice of gauge used to compute it -- only the
physical quantities, computed from the 1PI action (for scattering amplitudes,
this is the sum over tree graphs contributing to a process), need be invariant.

In fact, the gauge-fixing term in background field gauge is
not invariant under conformally rescaling the metric.  Consider the Yang-Mills theory on the cosmological
spacetime \FRWframe. We choose as background fields a vanishing gauge field,
$A_{\mu} = 0$, and a scalar field profile $\bar{\Phi}^i$ (with $i$ an index transforming as a
${\bf 6}$ under the $SO(6)$ R-symmetry)  corresponding to a brane at a fixed location on $S^5$ and
moving along the radial direction $r_p$ in the $M=0$ hyperbolic black hole spacetime.
Consider the background field gauge condition\foot{See also \JevickiQS.
To justify the last term, note that if we were to use
T-duality on a D3-brane at a point on $T^6$, this would become the
standard $D_{\bar{A}}^{\mu} A_{\mu} = 0$ background field gauge condition,
where $\bar{A}$ is the background gauge field.}
\eqn\gaugecond{
	G_p = \nabla^{\mu, (p)} A^{(p)}_{\mu} + i\sum_i [\bar{\Phi}^i_p,\delta\Phi^i_p] = 0\ ,
}
where $\delta\Phi^i$ is the fluctuating part of $\Phi^i$.
Due to the covariant derivative, this gauge condition is not invariant under Weyl transformations, so that:
\eqn\gaugeconds{
	G_p \to e^{2t/\ell} \left( \nabla^{\mu}A_{\mu}
		+ \frac{2}{\ell} A_t + i\sum_i [\bar{\Phi}^i, \delta\Phi^i]\right) = e^{2t/\ell} \tG\ .
}
Note that $\tG$
is not invariant under time reversal with respect to $\tdt$; this makes this gauge
condition a plausible choice for deriving \Poincaretilde\ as an effective action.

The possibility that \dbis,\Poincaretilde\ arise from the same Yang-Mills actions
computed in different gauges is suggested by the  related story for the
action of special conformal transformations on Yang-Mills theory in $\QR^4$.
The Yang-Mills theory is invariant under such transformations.
The DBI action for  D3-brane probe in AdS
is invariant under modified special conformal transformations, in which one
adds a field-dependent term to the transformation of the coordinates \refs{\MaldacenaRE}.
This arises from pulling the spacetime isometry back to the worldvolume fields of the D-brane.  Such
a modified transformation is {\it not}\ an invariance of the underlying Yang-Mills action.
However, the AdS/CFT correspondence states
that the DBI action {\it should}\ arise from the Yang-Mills theory out on the Coulomb branch,
after integrating out the W-bosons which become massive out on the Coulomb branch.
(This has been verified to leading order in the inverse distance along the Coulomb
branch \refs{\MaldacenaNX,\KiritsisTX}.)

In \JevickiQS, the authors pointed out that while the Yang-Mills action is indeed invariant
under the special conformal transformation without the field-dependent term,
the gauge fixing term for background field gauge is not invariant under special
conformal transformations.
In order to restore the gauge condition to its original form after a special conformal
transformation, a field-dependent gauge transformation is required. The authors
checked that to one loop, the combined action of the usual special conformal transformation and
the field-dependent gauge transformation on the low-energy effective action
is equivalent to the modified transformation of \MaldacenaRE, and is indeed a symmetry
of the effective action.

In \S4.2\ we introduced a gauge invariance -- worldvolume reparametrization
invariance~-- and showed that the experience of different
observers was related by a combination of field redefinition and change of
how this gauge freedom was fixed.  Here we have argued that
these descriptions can also be related by changing the $U(N)$ gauge fixing condition.
If our suggestion holds, there is a nontrivial relationship between the
$U(N)$ gauge invariance and probe worldvolume plus spacetime diffeomorphisms.
It would be very interesting to study this further.

\newsec{Comments about the singularity}

We have seen that the hyperbolic black hole \hyperBH\ has an equivalent description in terms of the Poincare patch. For describing evolution across the horizon, the latter description is clearly preferable since the natural time slices  cross the horizon smoothly. Furthermore, a conformal transformation
of this theory to the static cylinder removes the singularity in the boundary spacetime and replaces it with the physics of the scalars near the origin.  In particular, the action \Poincaretilde\ allows
us to  follow D-brane probes all the way to the singularity, at $\phi = 0$.
Although we cannot control all aspects of the physics in this regime, we can make the following qualitative comments.

On the static cylinder, the scalars experience a $-Tr \Phi^2$ potential due to the negative curvature. Naively, it would appear that one could send in an eigenvalue with large positive energy and it would sail over the top of the potential. From the bulk standpoint, this would correspond to a D-brane entering the black hole, crashing through the singularity and reemerging to the future.  This striking conclusion is not correct. As we have discussed, there are various effects which can slow the evolution of the scalar near the origin.

Consider first the formation of the black hole by a collapsing  shell of D3-branes.
Initially, the SYM scalars are diagonal, with the eigenvalues coming in from infinity. The off diagonal modes (W-bosons) are very massive and start in their ground state. As the eigenvalues approach zero, the off diagonal modes become light and are copiously produced, leading to a complicated,
strongly coupled, excited state. From the bulk standpoint, as the shell of D3-branes becomes smaller, open strings are excited. A brane cannot return to large radius unless the open strings attached to it decay. This traps the eigenvalues  near zero.  If N is strictly infinite, the eigenvalues will be trapped forever. This provides a microscopic description of the formation of a classical black hole. We expect that a similar story will describe a D-brane probe that is sent
in later: as it approaches the origin of moduli space, it will start to excite W-bosons and other
modes, and be trapped by them, by an effect similar to the scattering of D0-branes at
small impact parameter \refs{\DouglasYP}.

Consider a probe brane at constant $r_p$.  In the conformally transformed action,
the scalar satisfies $\phi = k e^{-\tdt/\ell}$. (This is an exact solution to the DBI action
\Poincaretilde.)
Since the mass of the W-boson is proportional to $\phi$, it will be produced when
 \eqn\production{{|\dot \phi| \over \phi^2} \sim 1}
This implies that $\phi \sim 1/\ell$. The bulk radial coordinate is related to $\phi $ by $\tir = \ell_s^2\phi \sim \ell_s^2/\ell $.  So  the W-bosons are produced  well inside the horizon. This is consistent with the fact that infalling observers do not see thermal radiation near the horizon.

  %\AshtekarCJ
\lref\AshtekarCJ{
  A.~Ashtekar and M.~Bojowald,
  ``Black hole evaporation: A paradigm,''
  Class.\ Quant.\ Grav.\  {\bf 22}, 3349 (2005)
  [arXiv:gr-qc/0504029].
  %%CITATION = CQGRD,22,3349;%%
}
%\BrownSD
\lref\BrownSD{
  A.~R.~Brown and E.~J.~Weinberg,
  ``Thermal derivation of the Coleman-De Luccia tunneling prescription,''
  Phys.\ Rev.\  D {\bf 76}, 064003 (2007)
  [arXiv:0706.1573 [hep-th]].
  %%CITATION = PHRVA,D76,064003;%%
}
%\BrownZZH
\lref\BrownZZH{
  A.~R.~Brown, S.~Sarangi, B.~Shlaer and A.~Weltman,
  ``A Wrinkle in Coleman - De Luccia,''
  Phys.\ Rev.\ Lett.\  {\bf 99}, 161601 (2007)
  [arXiv:0706.0485 [hep-th]].
  %%CITATION = PRLTA,99,161601;%%
}

If N is large but finite, the eigenvalues will be trapped for a long time, but eventually come out,
and the black hole will eventually decay.
We can estimate the time scale over which this occurs, in Schwarzschild coordinates.
Consider the $\mu = 0$ black hole.
The gauge theory on the static spacetime $\Sigma \times \QR$ is at finite temperature,
$T = \frac{1}{2\pi \ell}$.  The dynamics of a single brane is described by the action \dbis.
The static potential starts from $V = 0$ at the horizon, rises to a peak, and falls off as $-r^2$
at large $r$.  We will assume the D-branes start near the horizon; if we form
the black hole by a collapsing shell, the D-branes will slow down and
strongly fluctuate at this point, so that the state of the dual scalars should have appreciable
support near the horizon. For an $M=0$ black hole, the only scale in the integrand of \dbis\  is $\ell$,
so that the D-brane action scales as:
\eqn\euclact{
	S = \frac{\ell^4}{g_s (\alpha')^2} C = N C
}
where $C$ is a constant of order 1.\foot{For $\mu > 0$, $C$ will be replaced by a function
$C(r_+/\ell)$, where $r_+$ is the horizon radius.}  Therefore, we expect that the timescale for emitting
a single brane is of order $e^{C N}$, and the time scale for emitting all $N$ branes is of
order $N e^{CN}$. This gives a semiclassical process for the black hole
to radiate away its charge.  The time scale is much longer than the evaporation time for a black hole
in asymptotically flat space or for a ``small" ten-dimensional black hole in global AdS; on the other hand, it is much shorter than the Poincar\'e recurrence
time for a spherical black hole. The hyperbolic black hole has positive specific heat; without the ability to radiate away D-branes, the black hole would come into thermal equilibrium with its own Hawking radiation and not evaporate, as with planar black holes or large spherical black holes in AdS. The ability to radiate D-branes means that at late time, the black hole will decay (along with the cosmological constant). Since it takes an infinite time for the branes to reach infinity,  the asymptotic geometry remains $AdS_5\times S^5$. The late time behavior is best described by conformally transforming the static cylinder to the expanding cone. In this frame, the scalar eigenvalues are constant (since the outgoing D-branes stay at constant Poincare radius) and the excess energy redshifts to zero.

In previous discussions of the black hole information puzzle, it has often been suggested that violations of locality might occur inside the horizon. In our example, this is clearly happening near the singularity.  Away from the singularity, local position on a constant-$\tdt$
time slice is well defined since it corresponds to the position of the D-branes or the size of the scalar eigenvalues. However, as we have seen, near the singularity all off-diagonal elements of the scalar matrix become important, all of the eigenvalues interact strongly with these modes and with each other, the D-brane ``probes" are no longer good definitions of any geometry, and we expect any notion of
geometry to break down. This provides a mechanism for information to come out of the black hole.

What is the best description of the causal structure of the final semi-classical spacetime? It is {\it not} just a smoothing out of the spacetime near the singularity with the region outside the horizon unchanged.  If that were the case, asymptotic observers who see the shell collapse would not see any branes emerge. The branes would all end up in the second copy of the black hole geometry to the future of the singularity. Since the branes come out in finite time as seen by asymptotic observers, there are two possible Penrose diagrams (see Fig. 4). The figure on the left describes the standard picture of an evaporating black hole in AdS. Since it is clear from the gauge theory that the evolution enters a nongeometric phase (when all the nonabelian degrees of freedom are excited) and there is no loss of unitarity, we think the  figure on the right is a more accurate description of the spacetime. The notion of an event horizon clearly requires global causal relations which are not well defined in a spacetime with nongeometric regions. Thus a global event horizon is only well defined in the classical limit and does not have a meaningful quantum analog\foot{The idea that event horizons may only exist in the classical limit has been discussed before \AshtekarCJ\ using a different approach to quantum gravity.}. More local concepts such as trapped surfaces and apparent horizons will remain meaningful away from the regions of large curvature.

\ifig\evapbh{The diagram on the left is the standard picture of an evaporating black hole in AdS. We believe the diagram on the right is a better description of the physics since it is clear from the gauge theory that the evolution enters a nongeometric phase and remains  unitarity.
} {\epsfxsize3.0in\epsfbox{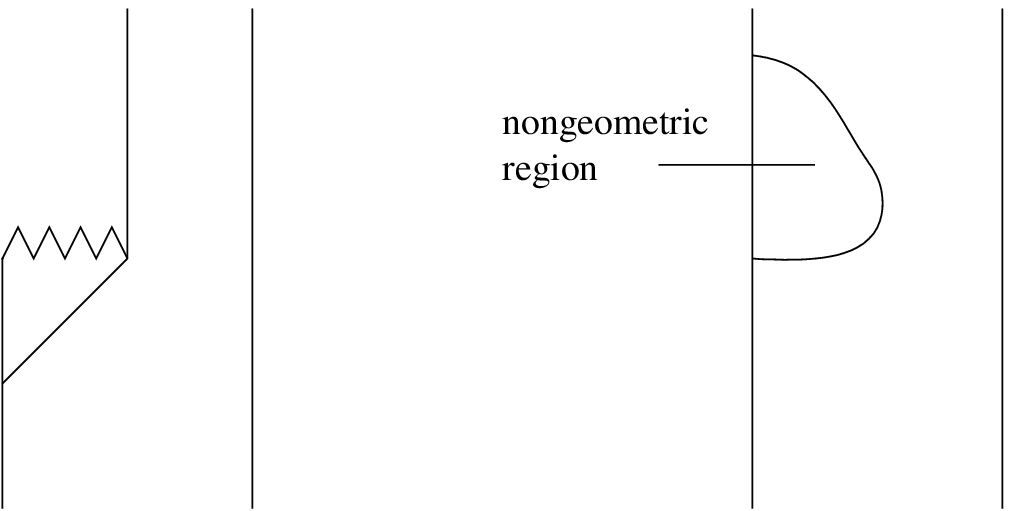}}

 Even though the singularity is resolved in our model, it is not a simple bounce. It takes a long time to pass through the singular region.  If a similar picture applied to cosmological singularities, it would have an important  implication.  It is usually assumed that superhorizon size perturbations propagate unaffected through a bounce.  This is because the bounce is assumed to happen quickly at one moment of time, so causality prevents any effects on large scale. This is not the case in our model. Due to the long time delay, there is no causality constraint. Indeed, we expect the W-bosons to eventually decay into inhomogeneous modes and change the spectrum of perturbations.

\newsec{Generalizations and open questions}

We have focussed on a particularly simple model of a black hole which is locally $AdS_5$ and used a ``dual" description in terms of Poincare coordinates to show how to describe physics inside the horizon. However the basic mechanism for going inside the horizon is much more general. Consider any  black hole in AdS. It can have arbitrary mass, and spherical, planar, or hyperbolic horizon. The usual Schwarzschild coordinates go bad at the horizon since the coordinate $t$ associated with time translations diverges there. One can introduce a good time coordinate associated with constant time surfaces that cross the horizon by a simple shift: $\tilde t = t + g(r)$ for some function $g(r)$. To study the motion of an infalling D3-brane one uses the fact that the radial position of the D-brane is associated with the eigenvalue of the super Yang-Mills scalars. The shift to a good time coordinate in the bulk thus corresponds to a field dependent time reparamaterization in the gauge theory. We can also follow D-branes falling into a
rotating black hole in AdS. In this case, good coordinates across the horizon are given by $\tilde \varphi = \varphi + h(r)$ as well as $\tilde t = t + g(r)$ for suitable functions $g,h$. In terms of the dual field theory, both the time and one of the
angular variables  must undergo field-dependent reparametrizations.

There are many directions for further work. One is to extend our proposal for infalling D-branes to a complete description of the physics inside the horizon in terms of the dual gauge theory. In the simple case of the hyperbolic black hole (with $ \mu=0$) that we have discussed for most of this paper, this is achieved by going to the Poincare patch description and using the gauge theory on the collapsing cone. But for a general black hole it is not clear what the analogous statement would be. It is possible that this is related to the discussion at the end of section 4, concerning the possible connection between our time reparameterization and Yang-Mills gauge transformations.

A second major open question is to gain better control over the singularity. The issue is no longer that our physical theory is breaking down near the singularity; it is clear that evolution in the gauge theory will continue for all time. The point is simply that  the dynamics involves all the light degrees of freedom in a complicated way. We believe that our general picture of the singularity will be applicable for all AdS black holes, but the details will differ. For example,  when strings propagate on negatively curved spacetimes of decreasing volume, it has been shown that there are many new light degrees of freedom \GreenTR, whereas this is not the case for positive curvature. For the hyperbolic black holes, compactifying the hyperbola (as we have done) leads to various complications such as introducing Wilson lines and a nonzero Casimir energy which can break the hyperbolic symmetry. Perhaps a first step toward understanding the physics near the singularity in this case is to consider the uncompactified theory.
One question which arises is the following. The breakdown of the DBI action in Schwarzschild coordinates corresponds to thermalization related to the presence of the horizon.
The breakdown of the DBI action in coordinates $\tdt,\tir$ occurs
near the singularity.  We expect a qualitative difference in the gauge theory between the physics of horizons (thermalization) and that of singularities,
since the spacetime near the horizon is weakly curved and classical, while the spacetime near the
singularity is strongly curved and the classical approximation is truly breaking down.  It would be interesting to find a way to characterize the distinction using our effective actions.\foot{We would like to thank S. Shenker for discussions on this point.}

A third question is to better understand the $M < 0$ black holes from the gauge theory.
These have rich causal structures, with multiple asymptotic regions, and inner and outer
horizons (which degenerate in the extremal limit).  While the singularities appear
timelike, the inner horizon is unstable to perturbations which generate initially null
singularities that turn over and become spacelike.  One possible attack on this class of black holes
is to study near-extremal black holes, using deviation from extremality as a small parameter.
More general questions are what the collapsing-shell spacetimes look like, and whether
there is a presentation of the gauge theory that reaches all of the asymptotic regions as well
as the interiors, or whether this is somehow cut off by the instability of the inner horizon.
Another interesting regime is for $M$ very small and negative; in this case, while the
gauge theory dynamics should not differ significantly from the $M = 0$ case, the
causal structure of the black holes at these energies is naively very different.

%End of paper
\vskip .5cm
\noindent{\bf Acknowledgements}

\nobreak We are grateful to B. Freivogel, M. Headrick, M. Kleban, J. Maldacena, R. Myers,  H. Schnitzer,
A. Sever,  S. Shenker and S. Trivedi for helpful discussions.  Part of this work was done while A.L. and E.S. were
attending the workshop ``Supersymmetry breaking and its mediation
in field theory and string theory" at the Aspen Center for Physics.
This work was completed while A.L. and E.S. were attending the ``Fundamental Aspects of Superstring
Theory" workshop at the Kavli Institute for Theoretical Physics at UC Santa Barbara.
E.S. is very grateful to the KITP and the Department of Physics at UCSB for
hospitality during several phases of this project. G.H. thanks the Department of Physics at Stanford University for hospitality.  A.L. is supported in part by DOE Grant
No.~DE-FG02-92ER40706, and by a DOE Outstanding Junior Investigator award.
E.S.  is supported by NSF grant PHY-0244728 and by the DOE under contract DE-AC03-76SF00515.
A.L. and E.S. are also supported by NSF grant NSF PHY05-51164. G.H. is supported by the NSF grant PHY-0555669.

\listrefs

\end